\documentclass[twocolumn]{aastex62}

\usepackage{booktabs}
\usepackage{longtable}
\usepackage{appendix}
\usepackage{amsmath}
\usepackage{amssymb}
\usepackage{graphicx}
\graphicspath{{figures/}}

\shorttitle{Mintz et al.}
\shortauthors{Mintz}

\begin{document}

\title{Constraining the Size of the Circumgalactic Medium Using the Transverse Autocorrelation Function of \ion{C}{4} Absorbers in Paired Quasar Spectra}

\author[0000-0002-9816-9300]{Abby Mintz}
\affil{Department of Astronomy, Yale University, New Haven, CT, United States}
\affil{Space Telescope Science Institute, Baltimore, MD, 21218, USA}
\affil{Maria Mitchell Observatory, Nantucket, MA, United States}

\author[0000-0002-9946-4731]{Marc Rafelski}
\affil{Space Telescope Science Institute, Baltimore, MD, 21218, USA}
\affil{Department of Physics \& Astronomy, Johns Hopkins University, Baltimore, MD 21218, USA}

\author[0000-0003-2973-0472]{Regina A. Jorgenson}
\affil{Maria Mitchell Observatory, Nantucket, MA, United States}

\author[0000-0001-6676-3842]{Michele Fumagalli}
\affil{Dipartimento di Fisica G. Occhialini, Universit\`a degli Studi di Milano Bicocca, Piazza della Scienza 3, 20126 Milano, Italy}

\author[0000-0002-6095-7627]{Rajeshwari Dutta}
\affil{Dipartimento di Fisica G. Occhialini, Universit\`a degli Studi di Milano Bicocca, Piazza della Scienza 3, 20126 Milano, Italy}

\author[0000-0001-9189-7818]{Crystal L. Martin}
\affil{Department of Physics, University of California, Santa Barbara, CA 93106, USA}

\author[0000-0003-0083-1157]{Elisabeta Lusso}
\affil{Dipartimento di Fisica e Astronomia, Universit\`a di Firenze, via G. Sansone 1, 50019 Sesto Fiorentino, Firenze, Italy}
\affil{INAF – Osservatorio Astrofisico di Arcetri, 50125 Florence, Italy}

\author[0000-0001-6248-1864]{Kate H. R. Rubin}
\affil{San Diego State University, Department of Astronomy, San Diego, CA 92182, USA}

\author[0000-0002-7893-1054]{John M. O'Meara}
\affil{W.M. Keck Observatory, 65-1120 Mamalahoa Highway, Kamuela, HI 96743}




\begin{abstract}

 The circumgalactic medium (CGM) plays a vital role in the formation and evolution of galaxies, acting as a lifeline between galaxies and the surrounding intergalactic medium (IGM). In this study we leverage a unique sample of quasar pairs to investigate the properties of the CGM with absorption line tomography. We present a new sample of medium resolution Keck/ESI, Magellan/MagE, and VLT/XSHOOTER spectra of 29 quasar pairs at redshift $2 <  z  < 3$. We supplement the sample with additional spectra of 32 pairs from the literature, creating a catalog of 61 quasar pairs with angular separations between 1.7\arcsec\ and 132.9\arcsec\ and projected physical separations ($r_\perp$) between 14 kpc and 887 kpc. We construct a catalog of 906 metal-line absorption doublets of \ion{C}{4} ($\lambda\lambda 1548, 1550$) with equivalent widths ranging from 6 m\AA $\leq W_{r, 1550} \leq 2053$ m\AA. The best fit linear model to the log-space equivalent width frequency distribution ($\log f(W_r) = m\log(W_{r}) + b$) of the sample yields coefficients of $m=-1.44\pm0.16$ and $b=-0.43\pm0.16$. To constrain the projected extent of \ion{C}{4}, we calculate the transverse autocorrelation function. The flattening of the autocorrelation function at low $r_\perp$ provides a lower limit for the coherence length of the metal enriched CGM – on the order of 200 $h^{-1}$ comoving kpc. This physical size constraint allows us to refine our understanding of the metals in the CGM, where the extent of \ion{C}{4} in the CGM depends on gas flows, feedback, timescale of metal injection and mixing, and the mass of the host galaxies.

\end{abstract}



\section{Introduction} \label{sec:intro}
Models of Big Bang nucleosynthesis and stellar nucleosynthesis provide well accepted constraints for the expected quantity of baryonic matter in the Universe \citep[][]{O'Meara2006, Spergal2006, Cooke:2018}. However, surveys of the low redshift Universe have revealed that the baryon content of galaxies is far below these expectations \citep{Fukugita2004}. Furthermore, surveys of the high redshift Universe ($z > 1.5$) have revealed significant metal enrichment in the intergalactic medium (IGM) as well as in the circumgalactic medium \citep[CGM,][]{Cowie1995, Tytler1995, Peroux2020}. These discoveries along with more recent studies \citep[][]{Peeples:2014, Werk:2014, Bordoloi:2014, Tumlinson:2017, Macquart:2020} indicate that there are significant deposits of diffuse, enriched gas surrounding galaxies. Cataloging these CGM deposits along with their redshift, metallicity, and column density \citep[][]{Rafelski:2012, Jorgenson:2013} and connecting those systems back to the associated galaxies \citep[][]{Tumlinson2013, Schroetter2016, Fumagalli:2017, Lofthouse2020, Rhodin:2021, Kaur:2021, Schroetter:2021} provides essential insight into galaxy evolution. 
The diffuse nature of the CGM makes it difficult to constrain the structure of its constituent gas clouds through direct imaging. Instead, the gas is typically studied in absorption to a bright background light source. Studies have used bright background galaxies \citep[][]{Cooke:2015, Peroux:2018, Mortensen:2021}, quasars \citep[][]{Wagoner:1967, Wolfe:1986, Chabanier:2021}, and gamma ray bursts \citep{Prochaska2009, Cuccchiara2015} as sources. However, these single sightline observations do not provide information about the characteristic spatial extent of the gas clouds, which constrains their origins, ranging from gas outflows, inflows, or reservoirs. 

In recent years, researchers have begun to examine samples of quasar pairs \citep[or galaxy pairs or galaxy - quasar pairs,][]{Rauch1999, Dodorico2006, Tytler:2009, Steidel2010, Martin2010, Chen:2014, Rubin:2015, Rubin:2018, Lee2018}. Paired background quasars broaden the analysis of the CGM by probing the gas at two different locations. This information can be used to measure the physical extent of the CGM via the transverse autocorrelation function, providing constraints for cosmological simulations \citep[][]{Rahmati:2013, Fumagalli2014, Peeples:2019}.

\begin{figure*}[t]
\begin{center}
\includegraphics[width=\linewidth,angle=0]{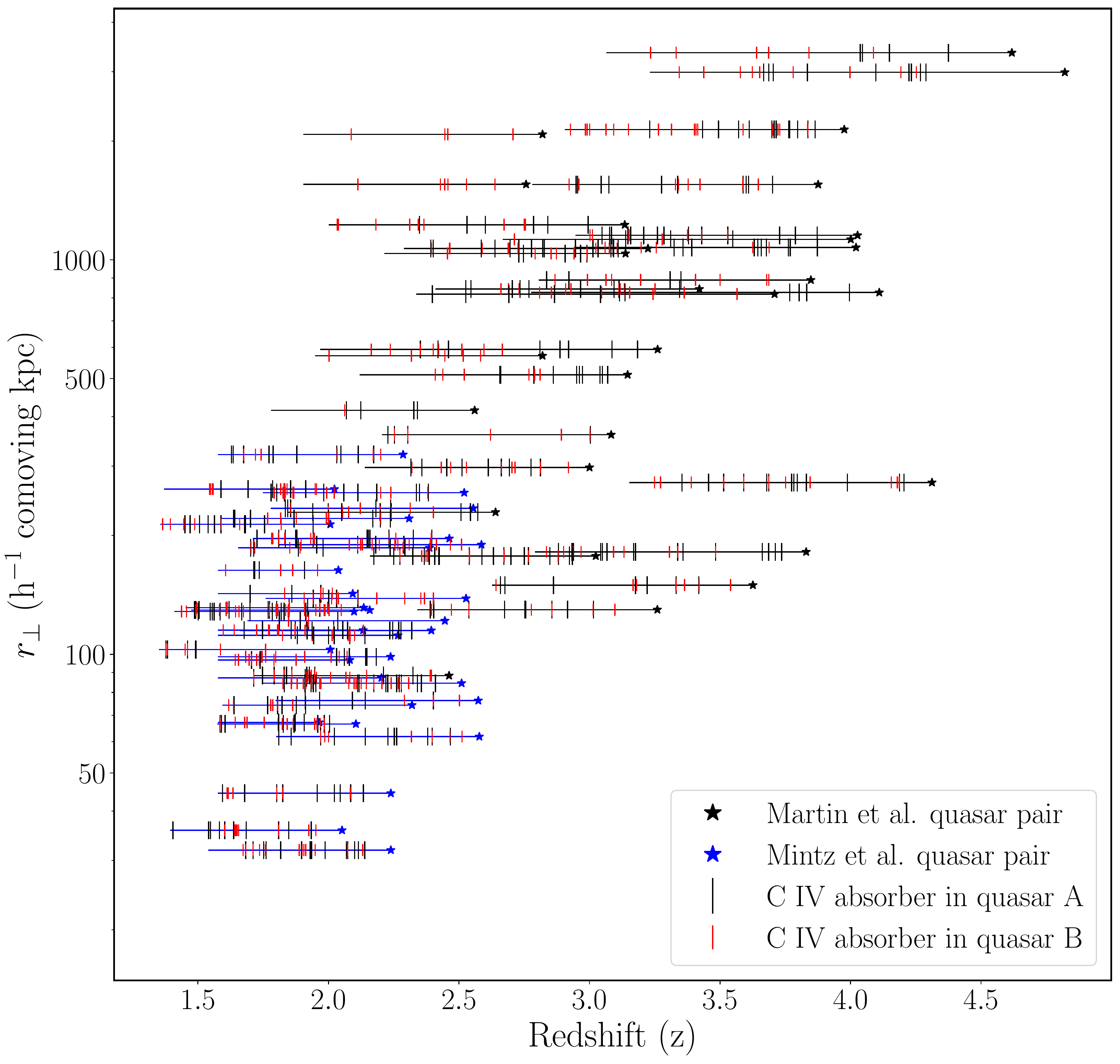}
\caption{Quasar pairs' search paths and detected \ion{C}{4} redshifts are shown as a function of pairs' projected separations. The search-path of each pair is plotted using the minimum of the two spectra's lower limits and the maximum of the two quasar's emission redshifts. The stars mark the higher emission redshift of the two quasars in a pair. The \ion{C}{4} absorbers found in the spectra of quasar A of a pair are shown in black and those found in quasar B are shown in red. The pairs from the Mintz et al. sample are plotted in blue while those from \citet{Martin2010} are plotted in black. The differences in the two samples including the $r_\perp$ ranges, redshift ranges, and lengths of search-paths are clearly visible in this plot. \label{fig:sightlines}}
\end{center}
\end{figure*}   

An effective tool for studying CGM and IGM enrichment is \ion{C}{4} absorption \citep[e.g.][]{Cooksey2010, Hasan2020}. The C IV absorption lines form a doublet with rest wavelengths of $\lambda\lambda$1548.20, 1550.77 \AA. The characteristic wavelength separation makes the doublet easily identifiable in spectra, even those with moderate resolution. The ratio of the transitions' oscillator strengths is 2:1, providing another avenue for verification. These features are both commonly found redward of Ly$\alpha$ emission and are redshifted into optical wavelengths for high redshifts and so are ideally positioned in wavelength space for the redshift range we investigate ($1.96<z<4.82$).  Since the pioneering work by \citet{Chen:2001}, it has been recognized that C IV absorption traces the gas around galaxies (i.e. CGM) with a high covering fraction out to $\sim$100-200 kpc and is kinematically consistent with residing in the same dark matter halo \citep{Adelberger:2005,Borthakur:2013, Bordoloi:2014, Turner:2017, Rudie:2019, Schroetter:2021}. 

In this work, we combine a new quasar pair sample, presented here for the first time, with additional data from the literature. Using the combined sample, we compile a catalog of \ion{C}{4} absorbers and use the paired nature of our sample to calculate the \ion{C}{4} transverse autocorrelation function to constrain the coherence length of the CGM as probed by this ion. Characterizing the spatial structure of CIV absorption constrains the origin of the gas, distinguishing between different models of gas flows and mapping the spatial distribution of gas around galaxies \citep{Chen:2014, Fumagalli2014}.

The paper is organized as follows. In Section~\ref{sec:data} we discuss the data set – a new sample of medium resolution spectra of 29 closely spaced quasar pairs. We supplement the sample with 3 and 29 additional quasar pairs from \citet{Rubin:2015} and \citet{Martin2010} respectively. In Section~\ref{sec:C4} we describe the \ion{C}{4} absorber doublet search, including details about the selection and evaluation of candidate absorbers. In Section~\ref{sec:sensitivity} we conduct a Monte Carlo simulation to assess the sensitivity and completeness of our doublet search. In Section~\ref{sec:distribution}, we calculate and compare the equivalent width frequency distributions for the Mintz et al. sample and the sample from \citet{Martin2010}. In Section~\ref{sec:autocor}, we calculate the transverse autocorrelation function, which describes the number of paired absorbers found in paired quasar spectra compared to random expectation. We then discuss our results, suggesting a lower limit for the physical size of the high redshift CGM traced by \ion{C}{4} and conclude with a summary in Section~\ref{sec:conclusions}. The following cosmology is used throughout: $H_0=70 \text{ km~s$^{-1}$~Mpc$^{-1}$}, \Omega_M = 0.3, \Omega_\Lambda = 0.7.$ 

\section{Observations and Data Reduction} \label{sec:data}

The quasar sample described in this paper is based on the sample from HST program 14127 (PI Fumagalli) that observed 47 quasar pairs at proper transverse separations of $r_\perp \leq 250$ kpc in the redshift interval $z\sim2.0-2.6$ with the WFC3/G280 grism. The program was designed to search for intervening pairs of Lyman Limit Systems (LLSs) that were followed up with higher resolution ground based spectroscopy to more precisely measure the redshifts of the LLSs and quasars as done in \citet{Lusso:2018}. The HST data sample was drawn from a compilation of quasar pairs with g$^*<21$ mag \citep{Hennawi:2006, Hennawi:2010, Findlay:2018}. The high resolution, ground-based spectroscopy forms the basis of the new sample of quasar pairs described below. The follow-up observations prioritized the brighter quasars. Six of the quasar spectra (three pairs) were previously published in \citet{Rubin:2015}, but because the targets are part of the HST sample we include them in our sample. The final sample consists of 64 quasar spectra – corresponding to 32 pairs.

\subsection{ESI}

The 14 quasar pairs observed with the Echellete Spectrograph and Imager \citep[ESI;][]{Sheinis:2002} at the Keck 2 telescope were previously presented in \citet{Lusso:2018}, who provide updated spectroscopic redshifts of the quasars. ESI is a fixed format echellete that covers the observed wavelength range $\lambda\approx$ 4000-11000 \AA. These quasar pairs were observed in January and April 2017 under programs PID 2016B\_N032E and 2017A\_N133E, with seeing of $\sim1$\arcsec. The 0.75\arcsec\ slit was used and both quasars were placed on the same slit when possible, while quasars with large separations were observed independently. The 0.75\arcsec\ slit corresponds to a resolution FWHM of $\sim44$ km~s$^{-1}$, or R$\approx$7000. When possible, we dithered along the slit to reduce fringing at the longest wavelengths ($\lambda>$8000 \AA). The ESI data were reduced as described in \citet{Rafelski:2012} using  the \verb|XIDL| package \verb|ESIRedux| \citep{Prochaska:2003} developed by J. X. Prochaska in IDL, and the spectra were extracted with optimal extraction.

\subsection{X-Shooter}

A sub-sample of 11 pairs included in this analysis has been observed with the X-SHOOTER spectrograph \citep{Vernet2011} mounted on the Very Large Telescope as part of the ESO programs PID 099.A$-$0018 and 0100.A$-$0084 (PI Fumagalli). 
X-SHOOTER is a medium-resolution, cross-dispersed echelle spectrograph able to collect data over a large wavelength range ($300-2500$~nm) split across three arms (UVB, VIS and NIR). For these observations, the instrument was configured with a 0.8~arcsec slit in the UVB arm and a 0.9~arcsec slit in the VIS and NIR arm, yielding a resolution of $\approx 6700$, 8900 and 5600. Observations were collected in ESO Period 99 and 100 in good weather conditions and seeing between $0.7-1$~arcsec. 
The UVB arm was reduced with the standard ESO pipeline \citep{Freudling2013} and the VIS and NIR arms were reduced with {\ttfamily PypeIt}\footnote{\url{https://pypeit.readthedocs.io/en/latest/}} \citep{pypeit:joss_pub}. First, the individual frames are processed for basic calibration (bias, dark and flat). Next, the wavelength calibrations are computed with the aid of arc frames, and the orders are identified and traced. Lastly, following sky subtraction, the spectra are then extracted and flux calibrated.

\subsection{MagE}

For 7 quasar pairs in this sample, we make use of data collected with the Magellan Echellette \citep[MagE,][]{Marshall2008} spectrograph at the Magellan Telescope. Similarly to ESI, MagE is a fix-format cross-dispersed spectrograph covering in one exposure the range $300-1000~$nm at a resolution of 4100 and 5800 for typically-used slits of 0.7 or 1 arcsec. Observations have been collected through the years (2008-2015) with good weather conditions.
Data have been reduced using the MASE pipeline \citep{Bochanski2009}, which follows a very similar workflow to the one of the ESIRedux package described above. Three of the quasar pairs' spectra were previously published in \cite{Rubin:2015}, as indicated in Table~\ref{tab:pairs}.  

\subsection{Complete Sample}

The complete new sample of quasar pairs described above is summarized in Table~\ref{tab:pairs} and will be referred to throughout as the Mintz et al. sample. It consists of 64 medium resolution quasar spectra corresponding to 32 quasar pairs, 3 of which were published previously in \citet{Rubin:2015}. The quasars have emission redshifts ranging from $1.96 \leq z \leq 2.64$ and the pairs have projected separations of $32  \leq r_\perp\ ( h^{-1}\ \text{kpc})\leq 320$ (comoving). Throughout this work, we use the quasar emission redshifts from \citet{Findlay:2018} who estimated an uncertainty of $\approx 1000$ km~s$^{-1}$, but \citet{Lusso:2018} has since calculated more precise measurements for most quasars in the sample. As our analysis deals exclusively with intervening systems, a slight difference in redshift would have no affect on our results.  

We supplement our dataset with additional Keck/ESI spectra of quasar pairs from \citet{Martin2010} provided by the authors. The sample from \citet{Martin2010} includes 55 quasar spectra corresponding to 29 quasar pairs (the sample includes one quasar triplet which provides 3 quasar pairs). The Martin et al. sample has similar, but slightly lower resolution than our sample due to the use of a wider 1\arcsec\ slit compared to our use of a .75\arcsec\ one. It is also different in a number of key aspects: it probes a higher redshift range of $2.46\leq z \leq 4.82$, it covers a higher $r_\perp$ range of $88 \leq r_\perp\ (h^{-1}\ \text{kpc})\leq 3350$ (comoving), and overall has a higher signal to noise ratio (SNR) with a median SNR per pixel of 18.5 compared to a median SNR per pixel for our sample of 11.5. The higher SNR can be attributed in part to the lower resolution. The differences in redshift and $r_\perp$ coverage are shown in Figure~\ref{fig:sightlines}. 
While \citet{Martin2010} performed their own doublet search, to obtain a maximally self consistent catalog, we combine the Martin et al. quasars with our sample and search for \ion{C}{4} absorbers in all of the spectra using the method described in Section~\ref{sec:C4}. We note that we do not include the sample by \citet{Tytler:2009} as it is at lower resolution and has a significantly larger pair separation with a median separation of 1 Mpc. 

Throughout this paper, unless otherwise explicitly noted, when we refer to the quasar pair sample we are referring to the combined sample of 119 quasar spectra or 61 quasar pairs. When we wish to refer to each of the subsamples separately, for simplicity we will reference them as Mintz et al. and Martin et al. \citep{Martin2010} respectively. 

\subsection{Continuum Fitting}\label{subsec:cont}

Each of the spectra in the Mintz et al. sample are continuum fitted using the \verb|lt_continuumfit| GUI from the \verb|linetools| package in python \citep{Prochaska2016}. The GUI provided an initial guess using spline knots and then the initial fits are adjusted manually. The spectra from \citet{Martin2010} were already continuum normalized by the authors. 

\section{\ion{C}{4} Doublet Search} \label{sec:C4}

Each of the quasar spectra are searched automatically for candidate \ion{C}{4} absorbers, identified based on the characteristic wavelength separation of the \ion{C}{4} doublet. All of the candidates are then evaluated based on specific characteristics of \ion{C}{4} systems and manually assessed to account for and correct blends between \ion{C}{4} and other absorption lines. This section details the \ion{C}{4} detection process and the compilation of the absorber catalog. The line search we adopt is based on the approaches of \citet{Cooksey2008}, \citet{Cooksey2010}, and \citet{Masribas2018} and further optimized for our data and methodology. 

\subsection{Absorption Feature Detection}\label{subsec:absorption}
The automated doublet search first identifies all absorption features in the continuum normalized spectra. The spectra are smoothed using a Gaussian convolution with width equal to the FWHM of the instrument as provided in Table~\ref{tab:pairs}. Absorption candidates are identified as groups of three or more adjacent pixels with $f_G\leq 1-\sigma_f$ where $f_G$ is the Gaussian convolved flux and $\sigma_f$ its uncertainty. Absorption features are then selected from the candidates if they have $W_{obs}\geq 3\sigma_{W_{obs}}$ where:
\begin{align}
W_{obs}&=\sum_{\lambda_l}^{\lambda_h}(1-f_{\lambda_i})\Delta\lambda_i
\end{align}
\noindent and
\begin{align}
\sigma^2_{W_{obs}}&=\sum_{\lambda_l}^{\lambda_h}\sigma^2_{\lambda_i}\Delta\lambda_i^2,
\end{align}

\noindent bounded by the feature's lower and upper wavelength limits: $\lambda_l$ and $\lambda_h$. The bounds are defined by the first and last contiguous pixels that satisfy the flux requirements as described above. For each accepted absorption feature we calculate its optical depth weighted \ion{C}{4} 1548 and \ion{C}{4} 1550 redshifts defined as:

\begin{align}
1+z_{abs}=\frac{\sum^{\lambda_h}_{\lambda_l}\lambda_i\ln\bigg(\frac{1}{f_{\lambda_i}}\bigg)}{\lambda_r\sum^{\lambda_h}_{\lambda_l}\ln\bigg(\frac{1}{f_{\lambda_i}}\bigg)}
\end{align}
\noindent with $\lambda_r = 1548.195$ \AA\ and $\lambda_r=1550.770$ \AA\ respectively \citet{Cooksey2008}. Both the 1548 and 1550 redshifts are calculated as each absorption feature is considered as a candidate for both the \ion{C}{4} 1548 and \ion{C}{4} 1550 line in the next step of the automatic doublet candidate detection. For the same reason, each absorption feature's rest-frame equivalent widths are calculated for both redshifts as:

\begin{align}
W_r&=\frac{1}{1+z_{abs}}W_{obs}
\end{align}
\noindent and
\begin{align}
\sigma^2_{W_r}&=\frac{1}{(1+z_{abs})^2}\sigma^2_{obs}.
\end{align}

\noindent The final absorption line catalog is used to identify \ion{C}{4} doublet candidates.

\startlongtable
\begin{deluxetable*}{cllllllclc}
\tablecaption{Quasar Sample\label{tab:pairs}}
\tablewidth{0pt}
\tablehead{
\colhead{Name}	&\colhead{RA}	&\colhead{Dec}
&\colhead{$z_{em}$} &\colhead{g mag}	& \colhead{S/N} & \colhead{FWHM} &\colhead{Inst}	&\colhead{$\theta$} &\colhead{$r_\perp$} \\[-0.3cm]
\colhead{}	&\colhead{(J2000)}	&\colhead{(J2000)}	&\colhead{}	&\colhead{}	&\colhead{}	&\colhead{km~s$^{-1}$}	&\colhead{}&\colhead{\arcsec}&\colhead{$h^{-1}$ comoving kpc}
}
\startdata
J0134+2430A	& 01:34:58.860	& +24:30:50.57	& 2.105	& 20.42	& 10.0	& 55.9	& ESI	& 3.69	& 66.55	 \\
J0134+2430B	& 01:34:59.018	& +24:30:47.57	& 2.095	& 19.85	& 15.6	& 55.9	& ESI	& 3.69	& 66.55	 \\
J0735+2957A	& 07:35:22.429	& +29:57:10.17	& 2.082	& 20.57	& 6.9	& 55.9	& ESI	& 5.39	& 96.71	 \\
J0735+2957B	& 07:35:22.555	& +29:57:04.99	& 2.065	& 20.42	& 9.6	& 55.9	& ESI	& 5.39	& 96.71	 \\
J0813+1014A	& 08:13:29.491	& +10:14:05.25	& 2.098	& 19.39	& 19.9	& 54.5	& MagE	& 7.14	& 128.32	 \\
J0813+1014B	& 08:13:29.708	& +10:14:11.62	& 2.071	& 20.09	& 15.1	& 54.5	& MagE	& 7.14	& 128.32	 \\
J0846+2709A	& 08:46:24.330	& +27:09:58.44	& 2.203	& 20.67	& 4.9	& 55.9	& ESI	& 4.71	& 87.13	 \\
J0846+2710B	& 08:46:24.505	& +27:10:02.45	& 2.195	& 20.89	& 6.5	& 55.9	& ESI	& 4.71	& 87.13	 \\
J0852+3500A	& 08:52:30.221	& +35:00:03.66	& 2.238	& 20.39	& 9.8	& 55.9	& ESI	& 5.28	& 98.51	 \\
J0852+3459B	& 08:52:30.532	& +35:00:00.03	& 2.235	& 19.98	& 10.3	& 55.9	& ESI	& 5.28	& 98.51	 \\
J0920+1311A\tablenotemark{a}	& 09:20:56.009	& +13:11:02.66	& 2.427	& 19.20	& 35.5	& 60.0	& MagE	& 6.23	& 121.58	 \\
J0920+1310B\tablenotemark{a}	& 09:20:56.237	& +13:10:57.42	& 2.446	& 19.29	& 26.3	& 60.0	& MagE	& 6.23	& 121.58	 \\
J0937+1509A	& 09:37:47.249	& +15:09:28.02	& 2.555	& 20.05	& 12.7	& 54.5	& MagE	& 11.75	& 234.45	 \\
J0937+1509B	& 09:37:47.409	& +15:09:39.54	& 2.541	& 19.93	& 14.5	& 54.5	& MagE	& 11.75	& 234.45	 \\
J1002+3531A	& 10:02:33.904	& +35:31:27.58	& 2.305	& 18.87	& 18.2	& 55.9	& ESI	& 3.91	& 74.29	 \\
J1002+3531B	& 10:02:34.211	& +35:31:28.68	& 2.320	& 20.04	& 6.2	& 55.9	& ESI	& 3.91	& 74.29	 \\
J1045+4041A	& 10:45:33.318	& +40:41:38.01	& 2.261	& 20.30	& 9.1	& 55.9	& ESI	& 17.03	& 320.87	 \\
J1045+4041B	& 10:45:33.544	& +40:41:21.15	& 2.286	& 20.25	& 9.0	& 55.9	& ESI	& 17.03	& 320.87	 \\
J1056-0059A\tablenotemark{a}	& 10:56:44.883	& -00:59:33.43	& 2.135	& 20.12	& 14.4	& 54.5	& MagE	& 7.21	& 131.20	 \\
J1056-0059B\tablenotemark{a}	& 10:56:45.248	& -00:59:38.11	& 2.126	& 20.96	& 8.7	& 54.5	& MagE	& 7.21	& 131.20	 \\
J1104+2907A	& 11:04:30.006	& +29:07:53.48	& 2.133	& 20.35	& 9.6	& 55.9	& ESI	& 6.33	& 115.07	 \\
J1104+2907B	& 11:04:30.348	& +29:07:49.02	& 2.127	& 20.48	& 11.4	& 55.9	& ESI	& 6.33	& 115.07	 \\
J1139+4143A	& 11:39:47.063	& +41:43:51.15	& 2.202	& 20.21	& 8.4	& 55.9	& ESI	& 2.39	& 44.41	 \\
J1139+4143B	& 11:39:47.259	& +41:43:52.10	& 2.239	& 19.98	& 11.5	& 55.9	& ESI	& 2.39	& 44.41	 \\
J1150+0453A	& 11:50:31.143	& +04:53:53.26	& 2.527	& 20.62	& 8.6	& 54.5	& MagE	& 6.97	& 138.51	 \\
J1150+0453B	& 11:50:31.543	& +04:53:56.85	& 2.517	& 20.52	& 27.0	& 54.5	& MagE	& 6.97	& 138.51	 \\
J1236+5220A	& 12:36:35.145	& +52:20:59.08	& 2.567	& 20.62	& 9.9	& 55.9	& ESI	& 3.08	& 61.82	 \\
J1236+5220B	& 12:36:35.422	& +52:20:57.33	& 2.578	& 20.50	& 10.7	& 55.9	& ESI	& 3.08	& 61.82	 \\
J1332+2523A	& 13:32:09.268	& +25:23:01.36	& 2.080	& 20.15	& 8.1	& 55.9	& ESI	& 7.92	& 142.42	 \\
J1332+2523B	& 13:32:09.690	& +25:23:06.83	& 2.093	& 20.11	& 6.2	& 55.9	& ESI	& 7.92	& 142.42	 \\
J1345+2625A	& 13:45:43.640	& +26:25:06.94	& 2.038	& 20.27	& 12.1	& 55.9	& ESI	& 9.22	& 163.35	 \\
J1345+2625B	& 13:45:44.316	& +26:25:05.35	& 2.016	& 19.96	& 14.0	& 55.9	& ESI	& 9.22	& 163.35	 \\
J1431+2705A	& 14:31:04.647	& +27:05:24.64	& 2.266	& 19.81	& 4.4	& 55.9	& ESI	& 5.94	& 111.65	 \\
J1431+2705B	& 14:31:04.977	& +27:05:28.63	& 2.263	& 20.26	& 6.2	& 55.9	& ESI	& 5.94	& 111.65	 \\
J1549+3136A	& 15:49:38.172	& +31:36:46.89	& 2.520	& 20.12	& 11.5	& 55.9	& ESI	& 12.96	& 256.85	 \\
J1549+3136B	& 15:49:38.496	& +31:36:34.61	& 2.502	& 19.29	& 21.1	& 55.9	& ESI	& 12.96	& 256.85	 \\
J1613+0808A\tablenotemark{a}	& 16:13:01.693	& +08:08:06.05	& 2.382	& 19.57	& 20.1	& 60.0	& MagE	& 9.64	& 186.12	 \\
J1613+0808B\tablenotemark{a}	& 16:13:02.033	& +08:08:14.26	& 2.387	& 18.94	& 32.9	& 60.0	& MagE	& 9.64	& 186.12	 \\
J1637+2636A	& 16:37:00.877	& +26:36:13.73	& 1.965	& 20.70	& 3.3	& 55.9	& ESI	& 3.86	& 67.14	 \\
J1637+2636B	& 16:37:00.925	& +26:36:09.92	& 1.961	& 19.39	& 10.4	& 55.9	& ESI	& 3.86	& 67.14	 \\
J2103+0646A	& 21:03:29.249	& +06:46:53.33	& 2.574	& 20.61	& 19.0	& 54.5	& MagE	& 3.81	& 76.31	 \\
J2103+0646B	& 21:03:29.372	& +06:46:49.99	& 2.565	& 20.20	& 12.7	& 54.5	& MagE	& 3.81	& 76.31	 \\
J1338+0010A	& 13:38:31.532	& +00:10:56.24	& 2.309	& 20.57	& 11.2	& 44.7	& X-SHOOTER	& 11.65	& 220.93	 \\
J1338+0011B	& 13:38:31.962	& +00:11:05.93	& 2.296	& 20.94	& 10.0	& 44.7	& X-SHOOTER	& 11.65	& 220.93	 \\
J1339+1310A\tablenotemark{b}	& 13:39:07.139	& +13:10:39.65	& 2.239	& 18.71	& 46.7	& 44.7	& X-SHOOTER	& 1.71	& 31.87	 \\
J1339+1310B\tablenotemark{b}	& 13:39:07.235	& +13:10:38.69	& 2.237	& 19.13	& 72.2	& 44.7	& X-SHOOTER	& 1.71	& 31.87	 \\
J1421+1630A	& 14:21:49.008	& +16:30:27.13	& 2.463	& 20.77	& 13.2	& 44.7	& X-SHOOTER	& 10.02	& 196.51	 \\
J1421+1630B	& 14:21:48.794	& +16:30:17.59	& 2.454	& 20.44	& 16.5	& 44.7	& X-SHOOTER	& 10.02	& 196.51	 \\
J1515+1511A\tablenotemark{b}	& 15:15:38.477	& +15:11:34.84	& 2.052	& 18.60	& 30.0	& 44.7	& X-SHOOTER	& 2.01	& 35.78	 \\
J1515+1511B\tablenotemark{b}	& 15:15:38.592	& +15:11:35.95	& 2.051	& 18.22	& 24.5	& 44.7	& X-SHOOTER	& 2.01	& 35.78	 \\
J2248+0307A	& 22:48:56.831	& +03:07:00.25	& 2.394	& 20.88	& 6.3	& 44.7	& X-SHOOTER	& 5.94	& 114.88	 \\
J2248+0307B	& 22:48:57.226	& +03:06:59.54	& 2.392	& 20.62	& 4.2	& 44.7	& X-SHOOTER	& 5.94	& 114.88	 \\
J2348+0057A	& 23:48:19.584	& +00:57:21.50	& 2.158	& 18.92	& 8.7	& 44.7	& X-SHOOTER	& 7.07	& 129.42	 \\
J2348+0057B	& 23:48:19.191	& +00:57:17.59	& 2.152	& 20.77	& 20.6	& 44.7	& X-SHOOTER	& 7.07	& 129.42	 \\
J0344+1015A	& 03:44:07.031	& +10:15:20.52	& 2.007	& 19.55	& 10.8	& 44.7	& X-SHOOTER	& 12.13	& 213.55	 \\
J0344+1015B	& 03:44:06.644	& +10:15:09.83	& 2.002	& 20.54	& 9.6	& 44.7	& X-SHOOTER	& 12.13	& 213.55	 \\
J1016+2224A	& 10:16:52.884	& +22:24:12.12	& 2.023	& 20.29	& 17.6	& 44.7	& X-SHOOTER	& 14.83	& 262.20	 \\
J1016+2224B	& 10:16:53.946	& +22:24:13.70	& 2.020	& 20.94	& 10.7	& 44.7	& X-SHOOTER	& 14.83	& 262.20	 \\
J2214+1326A	& 22:14:27.034	& +13:26:57.01	& 2.007	& 20.37	& 11.8	& 44.7	& X-SHOOTER	& 5.84	& 102.77	 \\
J2214+1326B	& 22:14:26.793	& +13:26:52.34	& 1.995	& 20.60	& 14.0	& 44.7	& X-SHOOTER	& 5.84	& 102.77	 \\
J2243-0613A	& 22:43:25.048	& -06:13:50.33	& 2.586	& 20.84	& 9.8	& 44.7	& X-SHOOTER	& 9.44	& 189.61	 \\
J2243-0613B	& 22:43:25.681	& -06:13:51.00	& 2.578	& 19.14	& 14.8	& 44.7	& X-SHOOTER	& 9.44	& 189.61	 \\
J2242+0558A	& 22:42:04.632	& +05:58:30.44	& 2.510	& 20.62	& 12.9	& 44.7	& X-SHOOTER	& 4.27	& 84.44	 \\
J2242+0558B	& 22:42:04.373	& +05:58:28.63	& 2.503	& 20.97	& 16.2	& 44.7	& X-SHOOTER	& 4.27	& 84.44	 \\
\enddata
\tablenotetext{a}{ Spectrum published in \citet{Rubin:2015}}
\tablenotetext{b}{ Gravitational lens}
\tablenotetext{}{\textbf{Note.} The new sample of quasar pairs and the three pairs from \citet{Rubin:2015}. $z_{em}$ is the emission redshift of the quasar, g mag the quasar magnitudes from \citet{Hennawi:2006, Hennawi:2010, Findlay:2018}, S/N the signal to noise ratio per pixel, FWHM the full width at half maximum of the instrument, Inst the instrument, $\theta$ the angular separation of the paired quasars, and $r_\perp$ the transverse separation in comoving units.}
\end{deluxetable*}

\begin{figure*}
\begin{center}
\includegraphics[width=\linewidth,angle=0]{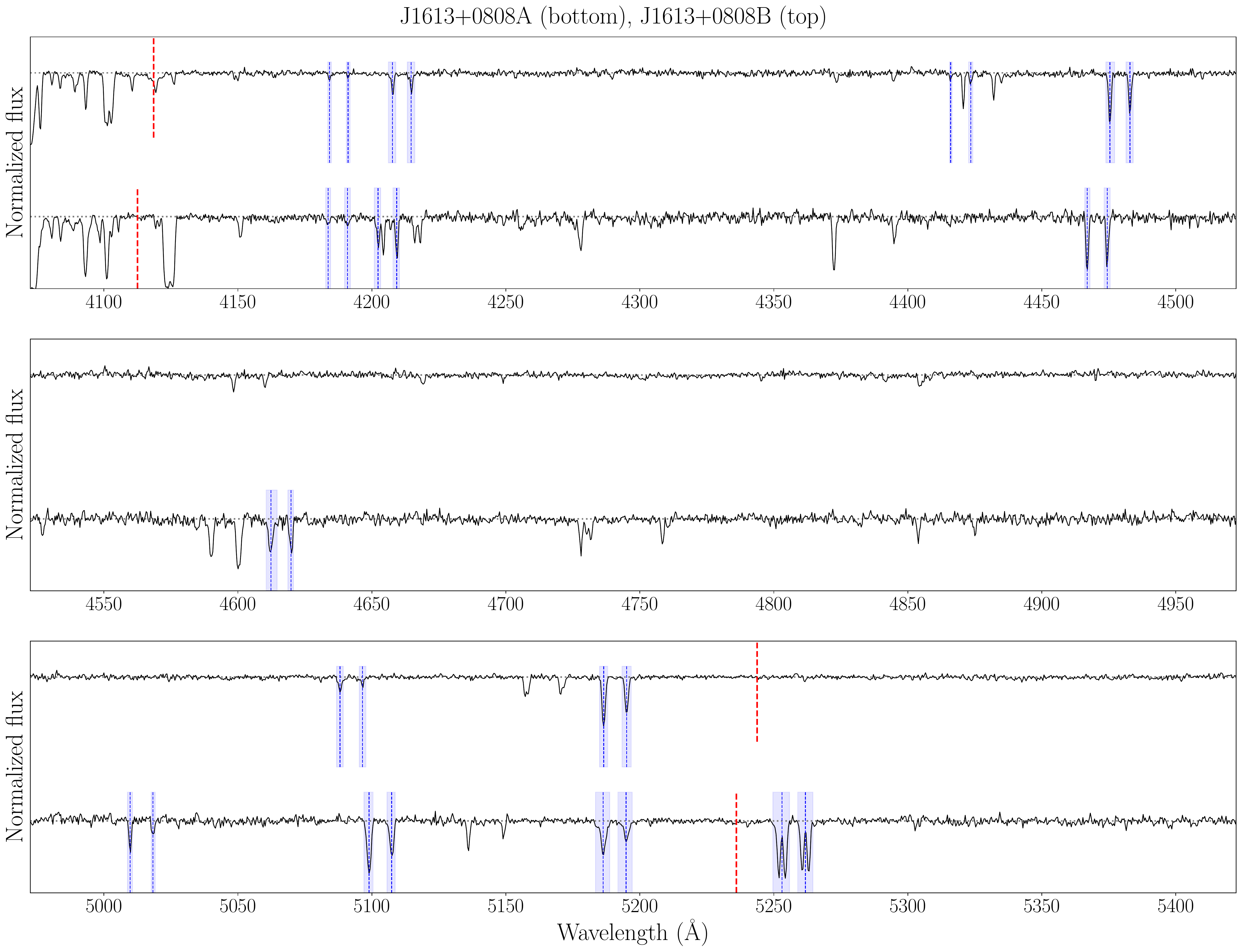}

\caption{An example of a \ion{C}{4} doublet search. (The complete set of doublet search plots are available in the figure set online.) The normalized flux is plotted in black, for the spectrum of J1613+0808A on the bottom and that of its pair, J1613+0808B, on top. The normalized continuum is shown as a dashed grey line. The lower limit of the search and the wavelength of \ion{C}{4} in the quasar's rest frame are marked by vertical, dashed red lines. The lower limit is determined by the location of the quasar's Ly$\alpha$ emission line and the upper limit extends 100\AA\ beyond the \ion{C}{4} wavelength at the quasar's redshift. This extension accounts for uncertainty in the quasars' emission redshifts and possible redshifting of \ion{C}{4} absorbers due to peculiar velocities. The likely \ion{C}{4} absorbers are indicated by the shaded blue regions which span the lines' equivalent width integration limits. The absorber redshift of the 1548 and 1550 \AA\ lines are shown with dotted, vertical blue lines. We see several examples of paired absorbers in this quasar pair – around 4190 and 5190 \AA. The absorbers around 4200, 4470, and 5100 \AA\ demonstrate the importance of the velocity width tolerance $\Delta v_{pair}$ discussed in Section~\ref{subsec:paircounts}. \label{fig:civsearch}}
\end{center}
\end{figure*}

\subsection{\ion{C}{4} Candidate Selection}\label{subsec:candidate}

The search for \ion{C}{4} doublets is limited to the region of the spectrum between the wavelength of the Ly$\alpha$ emission line (to avoid the Ly$\alpha$ forest which hinders line identification) and 100 \AA\ redward of the wavelength of \ion{C}{4} 1548 at the quasar's redshift. We extend the search slightly beyond the emission redshift to account for high uncertainty in the quasar redshifts and for possible peculiar velocities of the probed gas. We identify 37 systems above the emission redshift in a total of 33 quasar spectra, which are included in the catalog but flagged as proximate systems. In some cases, the wavelength coverage of a spectrum does not extend to the Ly$\alpha$ emission line or to the wavelength of \ion{C}{4} 1548 at the quasar's redshift, and in these cases the searchable path length is shorter. The search path also excludes regions with consistent contamination from skylines (i.e. 6860 \AA\ $<\lambda<$ 6930 \AA\ and 7590 \AA\ $ <\lambda<$ 7690 \AA). 

\ion{C}{4} absorber candidates are selected based only on the characteristic doublet separation. Each absorption feature detected as described in Section~\ref{subsec:absorption} is assumed to be the \ion{C}{4} 1548 line and the limits $\lambda_{l, 1550}$ and $\lambda_{h, 1550}$ of the potential \ion{C}{4} 1550 line are calculated as:

\begin{align}
\lambda_{l, 1550} = \lambda_{l,1548}\bigg(\frac{\lambda_{r, 1550}}{\lambda_{r, 1548}}\bigg)
\end{align}
\noindent and
\begin{align}
\lambda_{h, 1550} = \lambda_{h,1548}\bigg(\frac{\lambda_{r, 1550}}{\lambda_{r, 1548}}\bigg).
\end{align}

\noindent If a detected absorption feature falls in this range, the two features are selected as a \ion{C}{4} absorber candidate and added to the candidate catalog. If a single feature is sufficiently broad as to cover the \ion{C}{4} 1550 limits, it is included in the catalog and assessed manually as described in Section~\ref{subsec:evaluation}. This stage of candidate selection is intentionally designed to be generous in order to include systems that are exceedingly broad or blended with other lines.

\subsection{\ion{C}{4} Candidate Evaluation}\label{subsec:evaluation}
The \ion{C}{4} absorber candidates identified in Section~\ref{subsec:candidate} are then evaluated based on two additional characteristics of the \ion{C}{4} doublet: the equivalent width ratio and the velocity separation of the doublet lines. The equivalent width ratio and associated uncertainty are calculated as:

\begin{align}
    R_W &= \frac{W_{r,1548}}{W_{r, 1550}}
\end{align}
\noindent and
\begin{align}
    \sigma^2_{R_W} &= R^2_W\Bigg[\Bigg(\frac{\sigma_{W_{r, 1548}}}{W_{r,1548}}\Bigg)^2 + \Bigg(\frac{\sigma_{W_{r, 1550}}}{W_{r,1550}}\Bigg)^2\Bigg].
\end{align}

\noindent The \ion{C}{4} 1548 line is expected to be twice as strong as the \ion{C}{4} 1550 line in the unsaturated regime and so the \ion{C}{4} candidates that do not satisfy the following criteria are not included as likely \ion{C}{4} absorbers:

\begin{equation}
    1-3\sigma_{R_W} \leq R_W \leq 2+3\sigma_{R_W}.
\end{equation}

\noindent The lower limit of this criteria accounts for the possibility that the two lines are saturated and therefore could have nearly equal equivalent widths. We note that our $3\sigma$ tolerance is slightly higher than that used in \citet{Cooksey2010} and \citet{Masribas2018}. These studies did not perform a comprehensive visual inspection of the sightlines and so required more stringent selection criteria than our search. We inspect each sightline after the automatic search, manually correcting for blends and including multi-component systems which were not previously identified. The tolerance levels were chosen to maximize the number of true systems found, while minimizing false positives. 

The two lines of each doublet are expected to have the same redshift. To further select based on the lines' redshift difference, the velocity separation $\Delta v_\text{C~IV}$ is calculated as:
\begin{equation}
    \Delta v_\text{C~IV} = c\Bigg(\frac{z_{1550} - z_{1548}}{1+z_{1548}}\Bigg).
\end{equation}

\noindent Systems with $|\Delta v_\text{C~IV}| > 20 $ km~s$^{-1}$ are not included as likely \ion{C}{4} absorbers. Again, this tolerance is higher than that used in \citet{Cooksey2010} and \citet{Masribas2018} for the reasons described above.

In addition to the automatic evaluation based on $R_W$ and $\Delta v_\text{C~IV}$, the entire search-path is also visually assessed. Occasionally, the presence of blends or multiple velocity components leads to an incorrect evaluation of a candidate doublet. Such systems are manually flagged and the wavelength limits of their equivalent width integration are adjusted where appropriate. We link \ion{C}{4} absorber components separated by less than 250 km~s$^{-1}$ based on the distribution of damped Ly$\alpha$ \ion{C}{4} velocity widths reported in \citet{Prochaska2019}. An example of a full \ion{C}{4} doublet search is shown in Figure~\ref{fig:civsearch}.

We compile a catalog of 555 likely \ion{C}{4} absorbers in the Mintz et al. sample, from a candidate list of 2,567. As our primary area of interest in this study regards the intervening \ion{C}{4} absorbers, we flag any systems within 5000 km~s$^{-1}$ of the quasar redshift as proximate systems and remove them from subsequent analysis. We find 85 proximate systems, leaving 470 intervening \ion{C}{4} absorbers with redshifts in the range $1.36 \leq z \leq 2.58$ and a median redshift of 1.91. 

In the Martin et al. quasar sample, we find 524 likely \ion{C}{4} absorbers from a candidate list of 1224. Of the likely absorbers, 88 are proximate systems, leaving a total of 436 intervening absorbers with redshifts of $1.74 \leq z \leq 4.38$. The median redshift is 2.96. The acceptance fraction of likely absorbers from the candidate list is much higher in the Martin et al. sample due to the higher SNR of the quasars as compared to the Mintz et al. sample. This is discussed further in Section~\ref{sec:sensitivity}. We also note that our absorber catalog compiled from the Martin et al. sample is $\sim$~30\% larger than the catalog found in \citet{Martin2010}. Our methods for detecting \ion{C}{4} absorbers are not identical and so it is expected that the resulting catalogs would differ as well. For example, we do not require similar velocity structure in the two doublet lines as in \citet{Martin2010}, we do not separate multiple velocity components of absorbers, and we use a different linking length chosen as described above based on the results of \citet{Prochaska2019}. We refer the reader to \citet{Martin2010} for more details on their method. The variations in methodology and definition of \ion{C}{4} absorbers resulting in different sample sizes do not affect the results of the equivalent width distribution function or the autocorrelation as discussed in Sections~\ref{sec:distribution} and \ref{sec:autocor} as the same methodological differences are applied in the sensitivity assessment. In Section~\ref{sec:distribution} we find that the completeness-corrected distribution agrees with the \citet{Martin2010} results.

\begin{figure}[t]
\begin{center}
\includegraphics[width=\linewidth,angle=0]{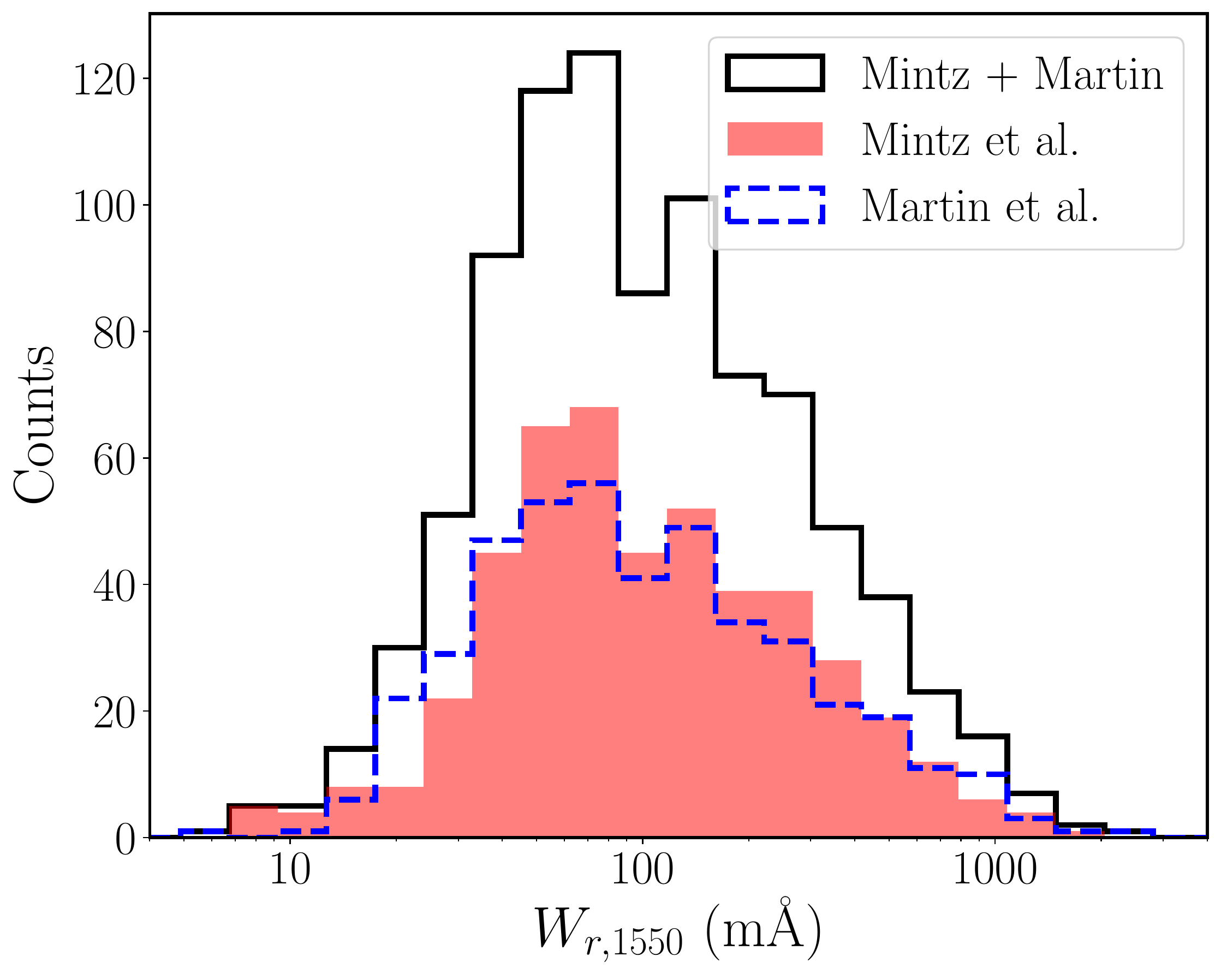}
\caption{A histogram of the $W_{r,1550}$ of the final \ion{C}{4} absorber catalog. The $W_{r,1550}$ distribution for absorbers from the Mintz et al. sample is shown in a shaded red histogram. The distribution for absorbers from the Martin et al. sample is shown with a dashed blue step histogram. The combined distribution is shown with a black step histogram. Despite the higher SNR of the Martin et al. sample, the two samples have similar lower $W_{r,1550}$ detection limits due to the higher resolution of the Mintz et al. sample. \label{fig:ewhist}}
\end{center}
\end{figure}

All together, we compile a catalog of 906 intervening \ion{C}{4} absorbers. In all analysis that follows, the \ion{C}{4} absorber catalog refers to this combined catalog described in Table~\ref{tab:absorbers}, whose $W_{r, 1550}$ distribution is shown in Figure~\ref{fig:ewhist}.

\begin{deluxetable*}{llllllllll}
\tablecaption{\ion{C}{4} Absorbers\label{tab:absorbers}}
\tablewidth{0pt}
\tablehead{
\colhead{Quasar}	&\colhead{z$_{abs}$}	&\colhead{$W_{r,1548}$}
&\colhead{$\sigma_{W_{r,1548}}$} &\colhead{$W_{r,1550}$}
&\colhead{$\sigma_{W_{r,1550}}$} & \colhead{$R_W$} &\colhead{$\sigma_{R_W}$}	&\colhead{$\delta v_{\text{C~IV}}$} &\colhead{Proximate}\\[+0.0cm]
\colhead{}	&\colhead{}	&\colhead{m\AA}	&\colhead{m\AA}	&\colhead{m\AA}	&\colhead{m\AA}	&\colhead{}	&\colhead{}&\colhead{km~s$^{-1}$}
}
\startdata
J0134+2430A	& 1.585	& 445	& 86	& 207	& 42	& 2.15	& 0.4	& 13.3	& False\\
J0134+2430A	& 1.590	& 406	& 72	& 248	& 60	& 1.64	& 0.4	& 6.9	& False\\
J0134+2430A	& 1.604	& 182	& 33	& 93	& 29	& 1.94	& 0.5	& -15.3	& False\\
J0134+2430A	& 1.827	& 113	& 21	& 173	& 22	& 0.66	& 0.2	& -5.3	& False\\
J0134+2430A	& 1.868	& 51	& 11	& 80	& 15	& 0.63	& 0.2	& 9.1	& False\\
J0134+2430A	& 1.959	& 76	& 19	& 111	& 20	& 0.68	& 0.3	& 13.7	& False\\
J0134+2430A	& 1.984	& 495	& 19	& 380	& 18	& 1.30	& 0.07	& -12.3	& False\\
J0134+2430A	& 2.004	& 68	& 13	& 68	& 12	& 1.00	& 0.3	& 4.1	& False\\
J0134+2430A	& 2.089	& 56	& 9	& 66	& 9	& 0.85	& 0.2	& -9.4	& True\\
J0134+2430A	& 2.094	& 32	& 6	& 31	& 6	& 1.05	& 0.3	& 4.1	& True\\
\enddata
\tablenotetext{}{\textbf{Note.} The \ion{C}{4} absorber catalog. $z_{abs}$ is the absorber redshift defined as the redshift of the $W_{r, 1548}$ line. $W_{r,1548}$, $\sigma_{W_{r,1548}}$, $W_{r,1550}$, $\sigma_{W_{r,1550}}$ are the rest frame equivalent widths and uncertainties of the 1548 and 1550 lines respectively. $R_W$ and $\sigma_{R_W}$ are the equivalent width ratio and its uncertainty. $\delta v_{\text{C~IV}}$ is the velocity difference between the 1548 and 1550 lines. The Proximate column indicates if the absorber is located within 5000 km~s$^{-1}$ of the quasar. (We show only the first 10 absorbers, but this table is available in its entirety in machine-readable form.)}
\end{deluxetable*}

\begin{figure*}[t]
\begin{center}
\includegraphics[width=\linewidth,angle=0]{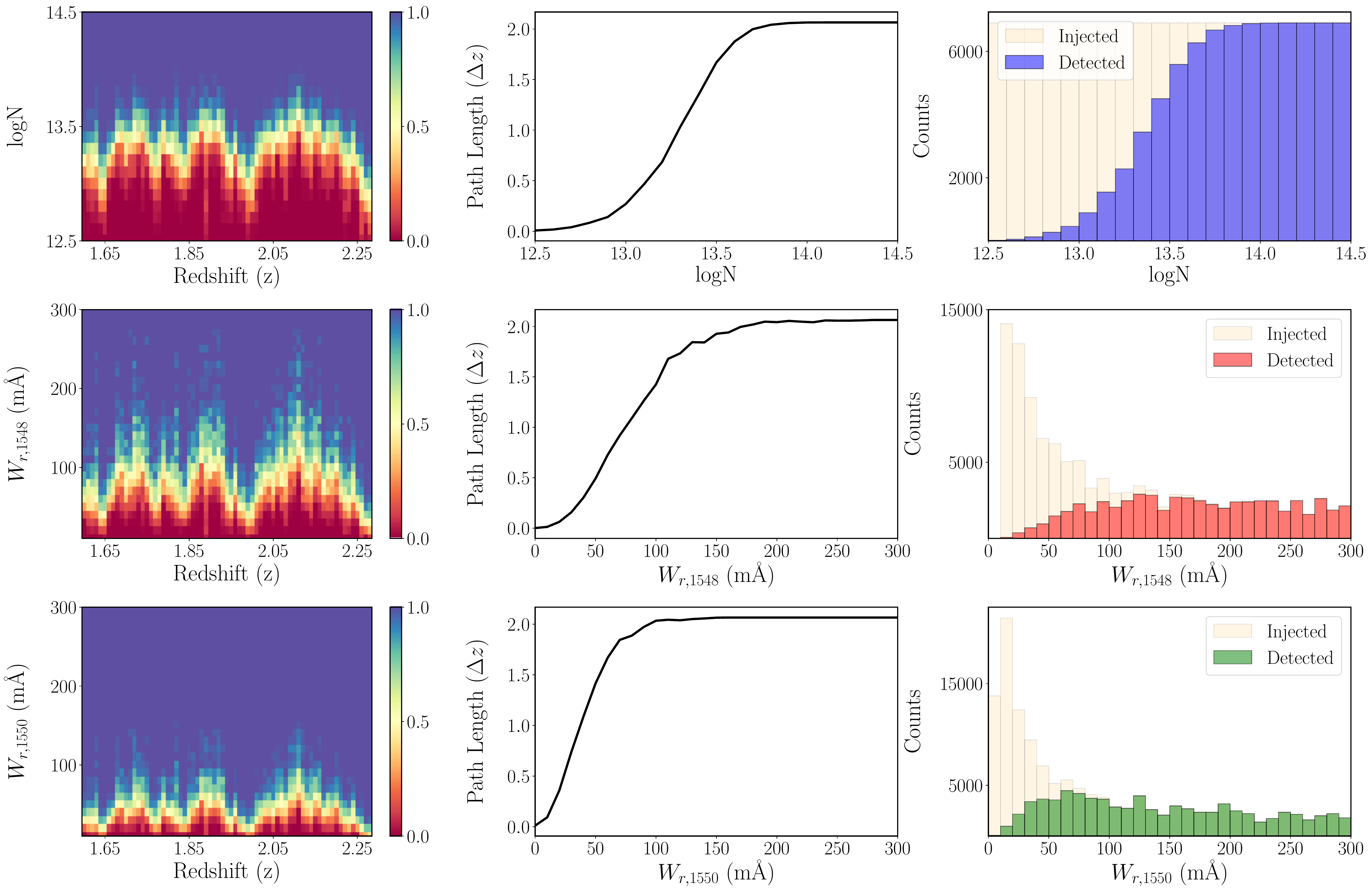}
\caption{An example of the completeness assessment for J100233.9+353127.5. The upper row shows the completeness in terms of logN, the middle row in terms of $W_{r,1548}$, and the bottom row in terms of $W_{r,1550}$. The leftmost column shows the completeness fraction as a function of redshift and system strength, with bluer values indicating higher completeness. The middle column shows the total searchable path length in units of $\Delta z$ as a function of logN or $W_{r}$. The rightmost column shows the number of injected versus detected systems in the completeness simulation. Note that the number of injected systems is constant in logN, but not in $W_{r}$ due to the injection method.\label{fig:completeness}}
\end{center}
\end{figure*}  

\begin{figure}[t]
\begin{center}
\includegraphics[width=\linewidth,angle=0]{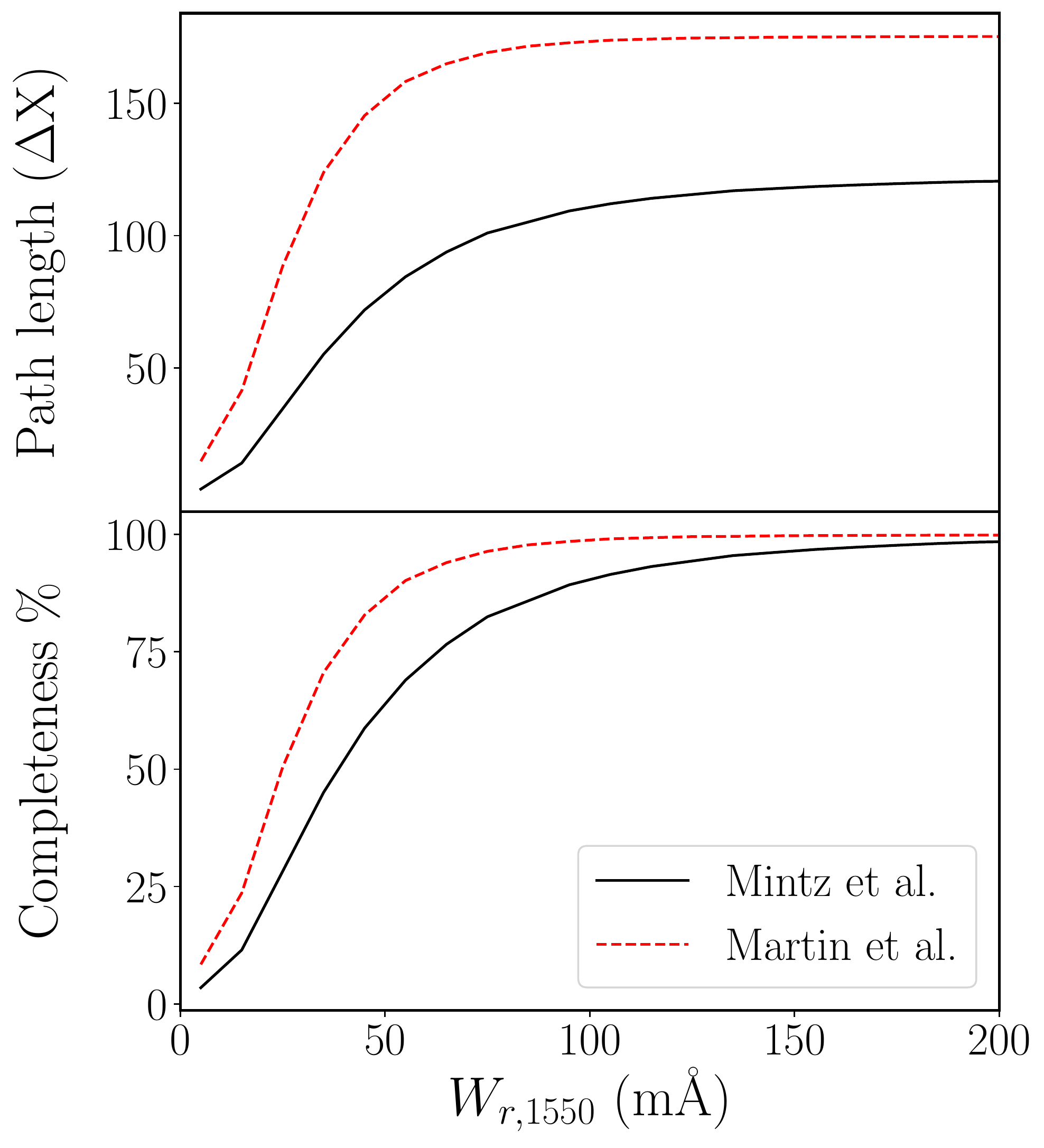}
\caption{Total searchable path in comoving units and completeness percentage are shown as a function of $W_{r,1550}$. The functions for the Mintz et al. sample are shown as black solid lines and those for the Martin et al. sample are shown as red dashed lines. Due to the higher emission redshifts of the quasars and the higher SNR of the spectra, the Martin et al. sample has a greater total searchable path and higher completeness for every $W_{r,1550}$. \label{fig:completenessfunc}}
\end{center}
\end{figure}

\section{Sensitivity and Completeness}\label{sec:sensitivity}

In order to conduct a proper analysis of our \ion{C}{4} absorber catalog, we must address the detection sensitivity of our spectra and the completeness of our absorber catalog. We follow the methods of \citet{Cooksey2010} and perform a Monte Carlo simulation to assess the sensitivity of the process described in Section~\ref{sec:C4}. First, each spectrum is cleaned of absorption features. The features are identified as described in Section~\ref{subsec:absorption} and replaced with flux randomly selected from the region adjacent to the feature so that the cleaned spectra accurately represent the noise of the original data. We choose to remove all absorption lines because we handled blends by eye in the construction of our absorber catalog.

Each spectrum is then injected with simulated \ion{C}{4} doublets. The redshifts and column densities of the doublets for a given spectrum are selected so that they fill a grid of equally spaced redshift bins of width 0.01 covering the search-path range and equally spaced log column density bins ranging from $12.5<\log N(\text{\ion{C}{4}})[\text{cm}^{-2}] <15.5$. For each spectrum, 100 simulated doublets are drawn from each redshift-column density bin. For example, in one such simulation, there are 100 doublets with $13.5<\log N(\text{\ion{C}{4}}) [\text{cm}^{-2}] <13.6$ and $2.00<z_{1548} <2.01$. The Doppler parameters of the systems are drawn randomly and range from 10 km~s$^{-1}$ to 35 km~s$^{-1}$, reflecting the width of the real data systems. Each doublet is also randomly assigned up to 5 velocity components with randomly selected fractions of the total column density and velocity offsets up to 50 km~s$^{-1}$. The equivalent width of each simulated system is calculated by integrating over the doublet's flux prior to adding the spectral noise.

To assess the detection sensitivity, each simulated doublet is injected into the cleaned spectrum, which is then run through the \ion{C}{4} absorber candidate selection process described in Section~\ref{sec:C4}. The candidates are not evaluated automatically as in Section~\ref{subsec:evaluation} based on Equations~10 and 11, as these are intended to exclude false positive detections – i.e. noise or other absorption lines that are not truly \ion{C}{4} doublets. Given that all of the absorption lines in the simulated spectra are by definition \ion{C}{4}, we do not apply these criteria. Furthermore, true systems with blends or multiple components may be discarded in the automatic evaluation. While we would recover these systems with our manual inspection in the true data, it is impossible to do so for the thousands of simulated spectra. Therefore, relaxing the selection slightly by omitting the automatic evaluation more accurately represents the conditions of the search on the actual data. 

To measure the searchable path as a function of column density and equivalent width, we first convert from redshift space to comoving absorption path length space (X) following \citet{Cooksey2010} using $\delta z = 0.01$ from the Monte Carlo simulation.
\begin{align}
    X(z) &= \frac{2}{3\Omega_M}\sqrt{\Omega_M(1+z)^3+\Omega_\Lambda},\\
    \delta X(z) &= X(z+0.5\delta z) - X(z-0.5\delta z).
\end{align}
\noindent We calculate the total searchable path for a given system strength as a weighted sum over the redshift bins, multiplying the width of the bin by the fraction of doublets of that strength detected in the simulation at that redshift. We define $C_i(\log N)$ to be this fraction in redshift bin $i$ for a system with column density of $\log N$ with equivalent definitions for $W_{r, 1548}$ and $W_{r, 1550}$. The total path length ($\Delta X$) as a function of column density (N) or equivalent width is then calculated as a sum over redshift bins $i$:

\begin{align}
    \Delta X(\log N) &= \sum\limits_{i}\delta X * C_{i}(\log N),\\
    \Delta X(W_{r,1548}) &= \sum\limits_{i}\delta X * C_{i}(W_{r, 1548}),\\
    \Delta X(W_{r,1550}) &= \sum\limits_{i}\delta X * C_{i}(W_{r, 1550}).
\end{align}

The total searchable path length function for each of the quasar pair samples is calculated as the sum of all individual searchable path length functions. The results of a single spectrum's sensitivity analysis is shown in Figure~\ref{fig:completeness} and the total completeness as a function of equivalent width is shown for both the Mintz et al. and the Martin et al. samples in Figure~\ref{fig:completenessfunc}. This completeness function further illustrates the difference in SNR between the two samples as discussed above. The Martin et al. sample is generally less noisy than the Mintz et al. sample due to the higher signal to noise (although at lower resolution), and so has higher completeness at all equivalent widths.

\begin{figure*}
\begin{center}
\includegraphics[width=0.48\linewidth,angle=0]{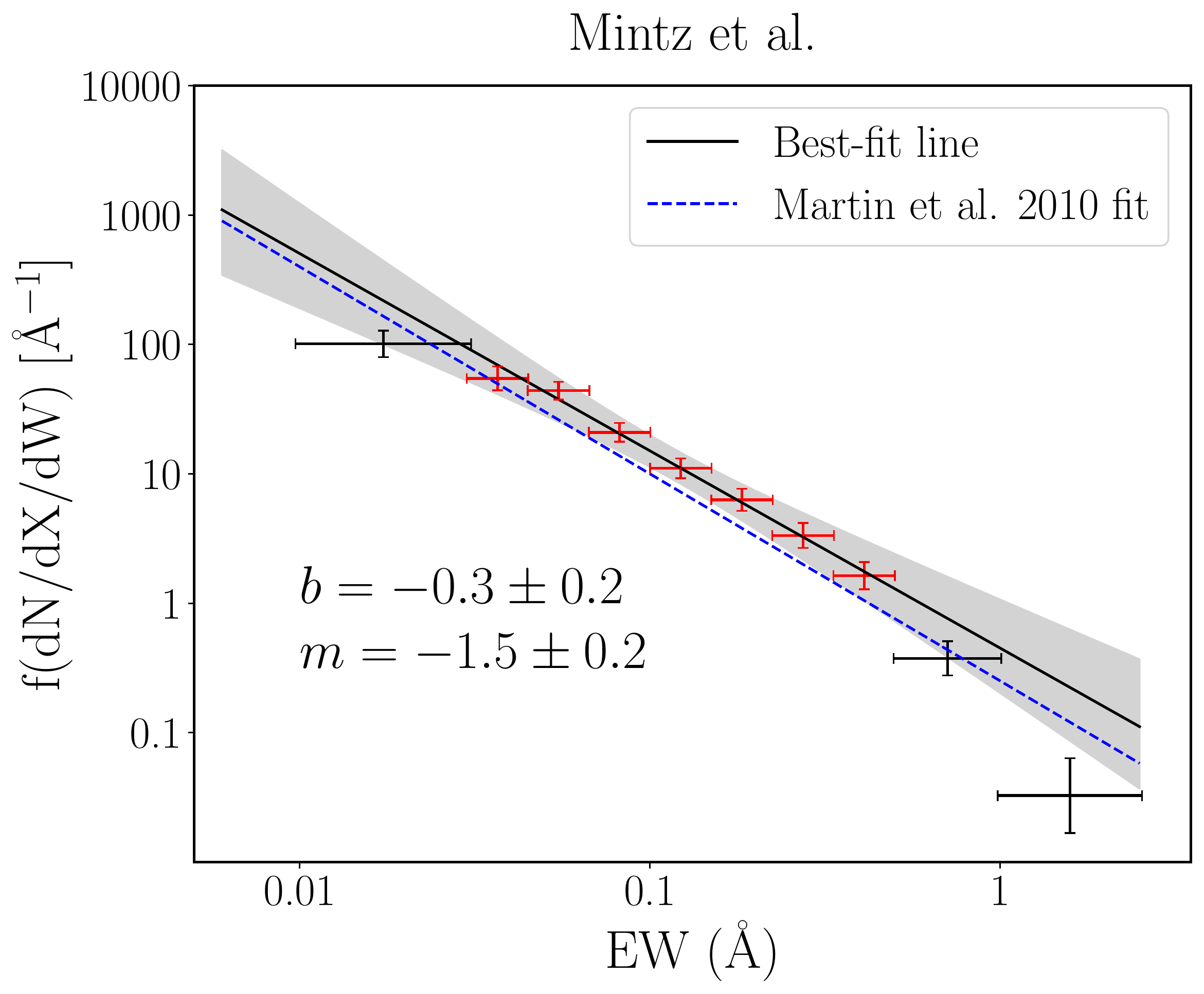}
\includegraphics[width=0.48\linewidth,angle=0]{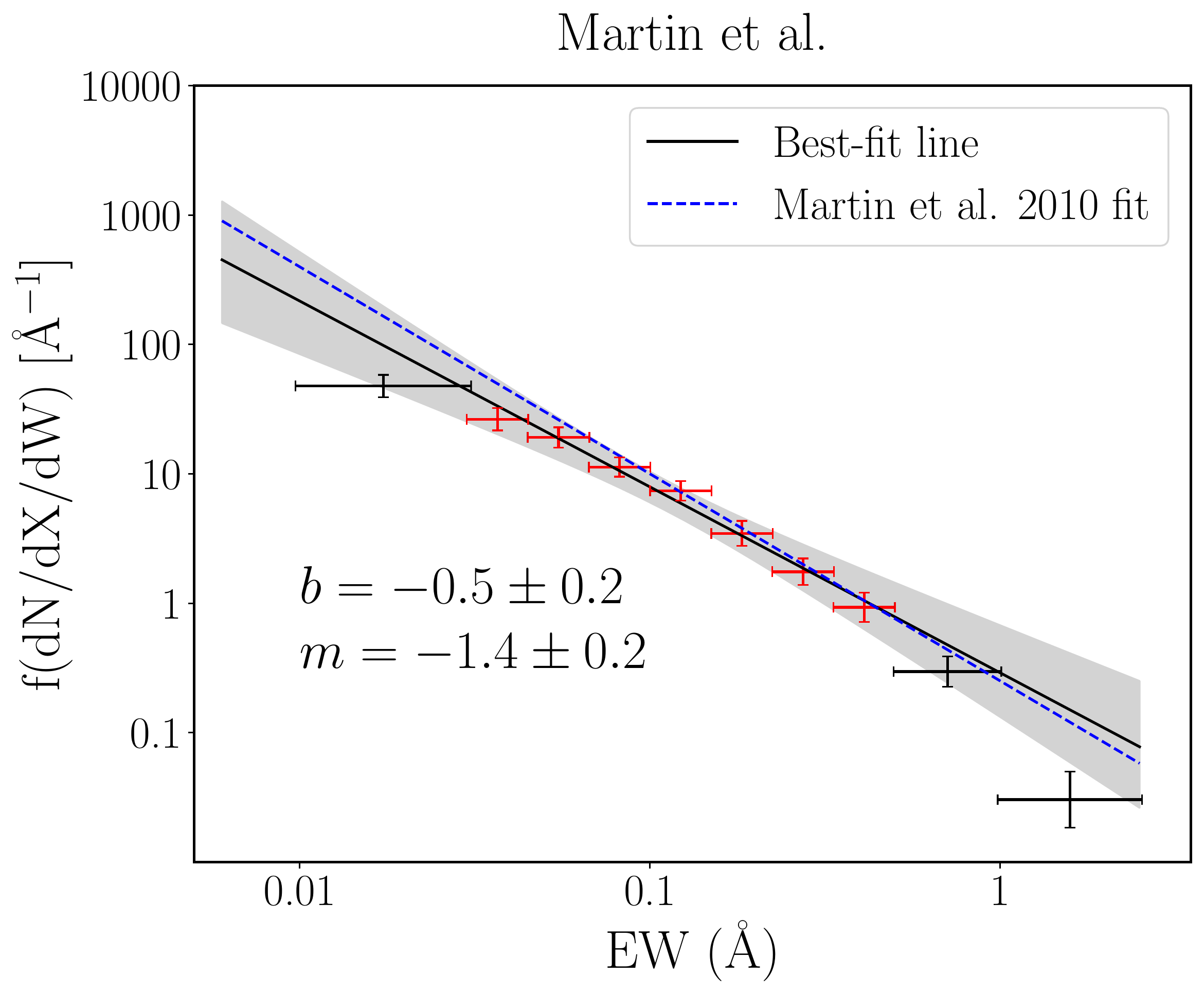}
\caption{Equivalent width frequency distributions for $W_{r,1550}$ for the absorbers found in the Mintz et al. sample (left) and those found in the Martin et al. sample (right). The black line shows the best fit line using only the bins plotted in red which fall between 30 and 500 m\AA, chosen to minimize bias from bins with few absorbers. The shaded grey region is the 2$\sigma$ confidence interval for the best fit line. The dashed blue line shows the best fit reported in \citet{Martin2010}, shown in both plots for comparison. Our calculated frequency distributions agree well with the published results despite slight differences in our methodology. \label{fig:ewdist}}
\end{center}
\end{figure*}

\section{Distribution Function}\label{sec:distribution}

Using the results from our Monte Carlo sensitivity assessment from Section~\ref{sec:sensitivity}, we calculate the equivalent width frequency distribution of \ion{C}{4} 1550 for the Mintz et al. sample and the Martin et al. sample, which is the number of systems per unit redshift path per unit rest equivalent width. The distribution function represents a nearly comprehensive summary of the absorption line survey. Following \citet{Martin2010}, we perform our analysis in terms of equivalent width, and not column density, as converting saturated systems' equivalent widths only provides a lower limit on column density. Additionally, to be conservative we use the equivalent width of the 1550 line instead of the 1548 line because it is the weaker of the two and detection of a doublet therefore depends on the detection of the 1550 line. 

The equivalent width distribution $f(W_r)$ is calculated following the method of \citet{Cooksey2010}:
\begin{equation}
    f(W_r) = \frac{\Delta \mathcal{N}}{\Delta W_r \Delta X(W_r)},
\end{equation}

\noindent where $\Delta\mathcal{N}$ is the number of absorbers per equivalent width bin, $\Delta W_r$ is the width of the equivalent width bin, and $\Delta X(W_r)$ is the searchable path length sensitive to absorbers of equivalent width $W_r$ calculated as described in Section~\ref{sec:sensitivity}. We calculate the uncertainty in the distribution function using the 84.1 \% Poisson confidence limits (corresponding to 1$\sigma$ for Gaussian statistics) for $\Delta \mathcal{N}$. 

It is well established that the distribution function can be described by a power law \citep{Cooksey2010, Martin2010}. Therefore, we use a Markov Chain Monte Carlo sampler to fit a linear model to our equivalent width distributions in log-log space using the width of the equivalent width bins as the uncertainity in the x-direction. For the fit, we use only the $f(W_r)$ bins with $W_r$ between 30 and 500 m\AA\ to avoid bias due to incompleteness and to compare with the results in \citet{Martin2010}. For the Mintz et al. sample, we obtain parameters for the fit to $\log f(W_r) = m\log(W_r) + b$ of $m=-1.5 \pm 0.2$ and $b=-0.3\pm 0.2$ for equivalent width in \AA\ and $b=4.1\pm 0.4$ for equivalent width in m\AA. For the Martin et al. sample, we found $m=-1.4 \pm 0.2$ and $b=-0.5\pm 0.2$ for equivalent width in \AA\ and $b=3.7\pm 0.4$ for equivalent width in m\AA. For both samples together, we found $m=-1.44 \pm 0.16$ and $b=-0.43\pm 0.16$ for equivalent width in \AA\ and $b=3.9\pm 0.3$ for equivalent width in m\AA. These fit values agree with those reported in \citet{Martin2010}, providing additional support for the consistency of our methodologies despite slight differences in our absorber detection strategies and catalogs. We show the distribution functions and their best fits in Figure~\ref{fig:ewdist}. We note that we do not expect the distribution function to be biased by our paired sightlines as systems found at similar redshifts in both sightlines of a pair do not necessarily have similar strengths. To confirm, we repeated the analysis described above for half of the sample, excluding one sightline of every pair, and obtained consistent results.

\citet{Cooksey2013} and \citet{Hasan2020} studied the evolution of the \ion{C}{4} distribution function with redshift using large samples of single-sightline background quasars. Their studies used \ion{C}{4} catalogs with thousands of absorbers with redshifts $1.0<z<4.75$. \citet{Cooksey2013} used low resolution SDSS spectra while \citet{Hasan2020} worked with high resolution spectra from Keck/HIRES and VLT/UVES. They both found evidence of higher \ion{C}{4} abundance at lower redshifts, which corresponds to a larger value for the intercept of the linear model. While slightly overlapping in redshift space, the Martin et al. sample is generally at a higher redshift than the Mintz et al. sample, with a median \ion{C}{4} absorption redshift of 2.96 compared to that of the Mintz et al. sample of 1.91. Based on \citet{Cooksey2013} and \citet{Hasan2020}, we would expect the Mintz et al. sample to have a larger intercept than the Martin et al. sample. We find that the slopes and intercepts of our best-fit lines to the Mintz et al. and the Martin et al. samples are statistically indistinguishable from one another, but attribute this to the small sample size. Our quasar-pair data set is not well suited for a thorough investigation of the redshift evolution of \ion{C}{4} enrichment, so we cannot comment confidently on agreement or disagreement with predictions. Paired quasars are not necessary for this type of analysis and are vastly more rare than unpaired quasars.

\begin{figure}
\begin{center}
\includegraphics[width=\linewidth,angle=0]{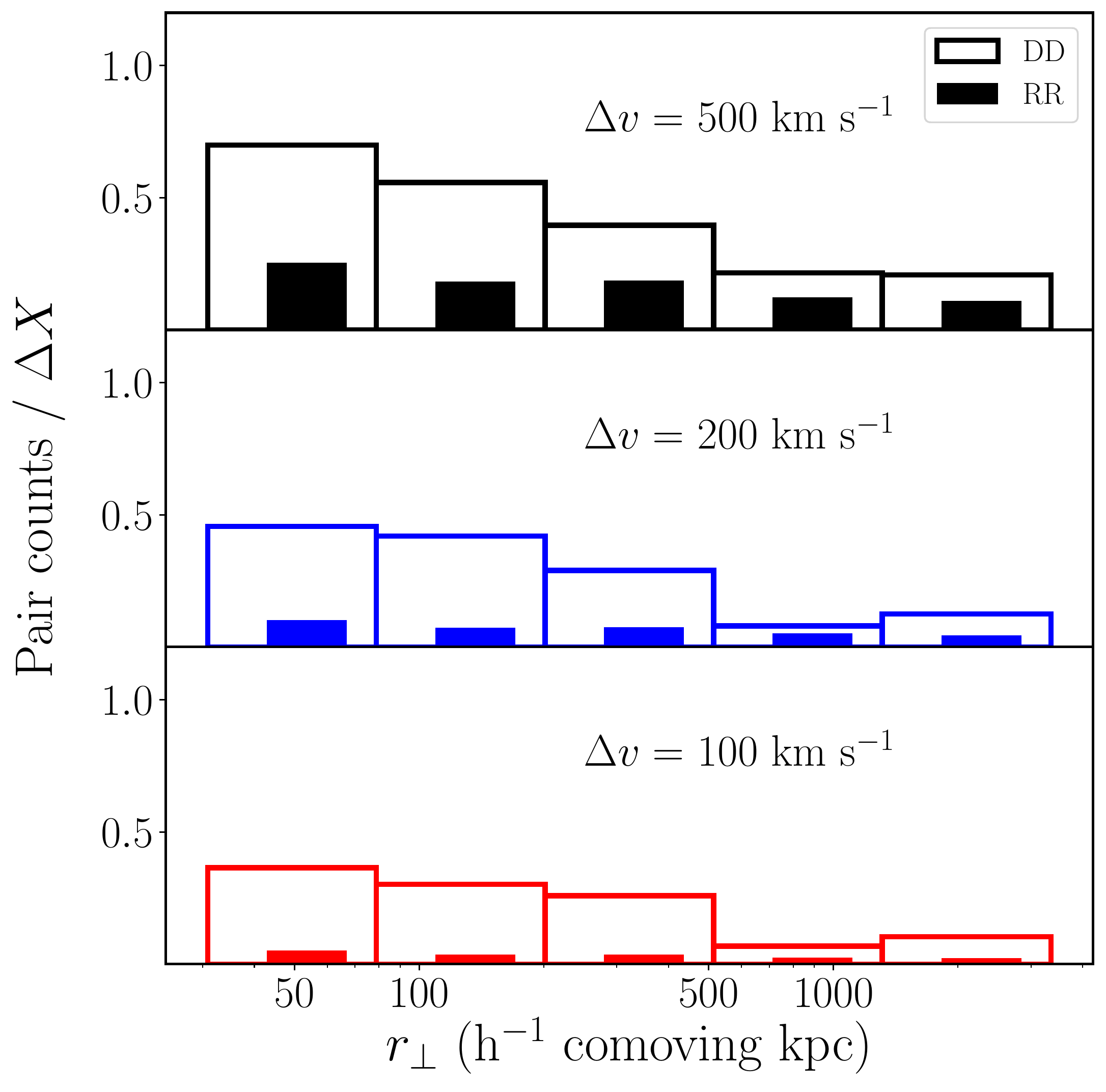}
\caption{A histogram of data-data (DD) and random-random (RR) counts for absorbers in the combined sample as a function of projected separation, normalized by total searchable path length per bin. The three rows show pair counts for velocity widths of $\Delta v_{pair} =$ 500, 200, and 100 km~s$^{-1}$. The empty bars show the the DD pair counts and the solid bars show the RR pair counts. There are significantly more DD pairs than RR pairs for all values of $r_\perp$ and $\Delta v_{pair}$. The number of pairs per bin increases as $\Delta v_{pair}$ increases, demonstrating the impact of the velocity width tolerance. We also see that the number of DD counts generally increases with decreasing $r_\perp$ but begins to flatten at $\approx$~200 $h^{-1}$ comoving kpc. \label{fig:achist}}
\end{center}
\end{figure}

\begin{figure}
\begin{center}
\includegraphics[width=\linewidth,angle=0]{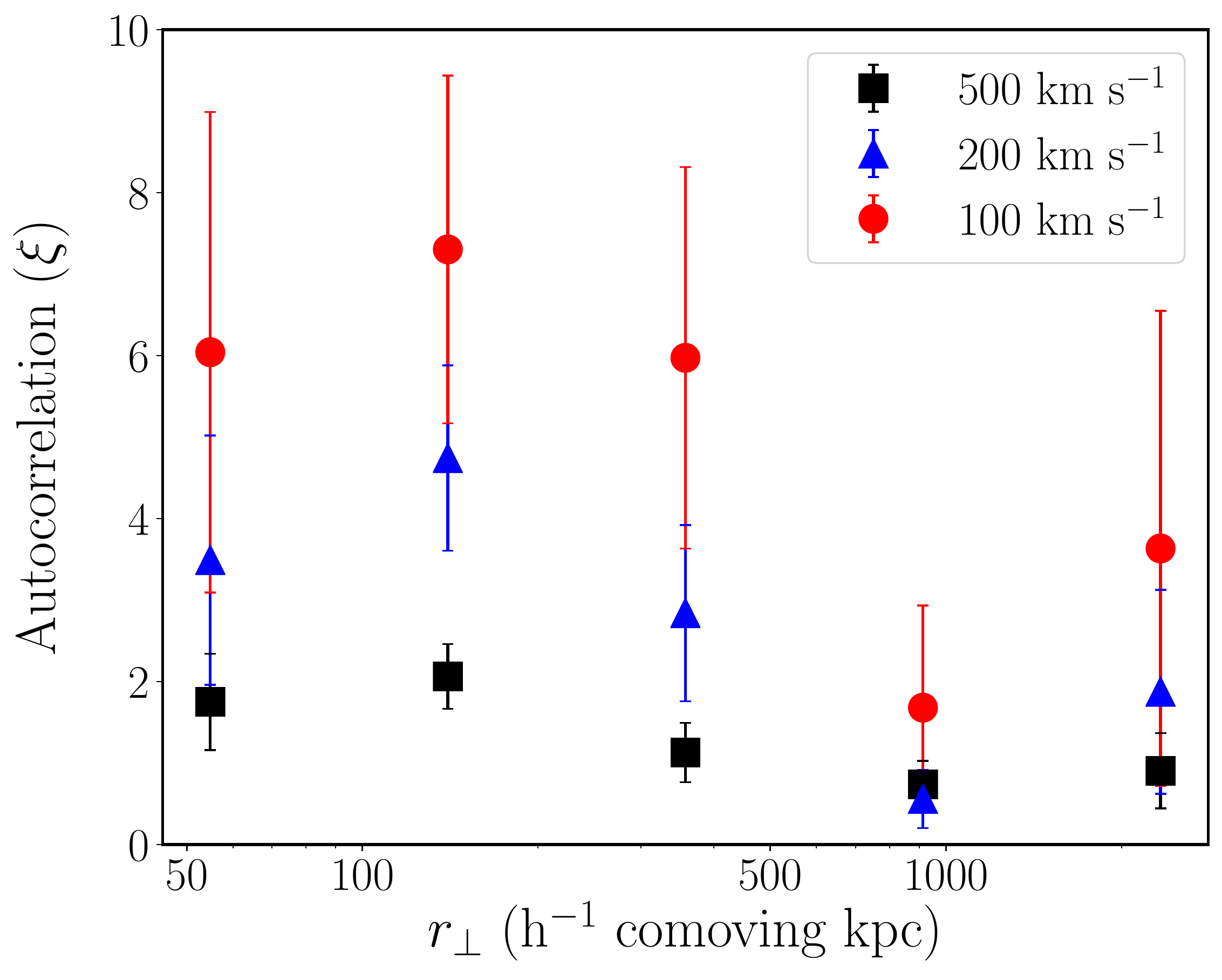}
\caption{The transverse autocorrelation function $\xi = \frac{DD}{RR}-1$ for the combined sample plotted as a function of perpendicular separation of the quasars pairs for three values of $\Delta v_{pair}$. As expected, smaller values of $\Delta v_{pair}$ yield higher values of $\xi$, indicating higher significance of paired absorbers. Following the trend shown in Figure~\ref{fig:achist}, $\xi$ generally increases with decreasing separation. The function begins to level off at $\approx$~200 $h^{-1}$ comoving kpc. This indicates that the structures we are probing are no smaller than $\approx$ 200 $h^{-1}$ comoving kpc. \ \label{fig:ac}}
\end{center}
\end{figure}

\section{Transverse Autocorrelation Function}\label{sec:autocor}

So far in this work, we have compiled a catalog of high redshift \ion{C}{4} absorbers and computed the equivalent width distributions of our sample. We have not yet utilized the unique paired nature of our quasar sample. As described in Section~\ref{sec:intro}, the transverse autocorrelation of absorbers in closely-spaced quasar pairs provides additional insight into the spatial distribution of metals in the CGM and yields constraints on the coherence length of the \ion{C}{4}, which is related to the physical size of the CGM as traced by this ion \citep{Adelberger:2005, Bordoloi:2014, Turner:2017, Rudie:2019, Schroetter:2021}. In this section, we calculate the transverse autocorrelation function for our sample following the methods of \citet{Hennawi2007}, \citet{Martin2010}, and \citet{Fumagalli2014}.

\subsection{Data absorber pair counts}\label{subsec:paircounts}

A careful definition of paired absorbers is essential for computing and understanding the implications of the autocorrelation function. An absorber pair is a group of two absorbers found at similar redshifts, but in different sightlines of a quasar pair. There are several examples of paired absorbers in Figure~\ref{fig:civsearch} at $\lambda\lambda 4190$ and 5190 \AA. 

Whether or not two absorbers in paired quasar spectra are considered an absorber pair is dependent on the velocity width tolerance. A smaller tolerance will lead to a smaller number of paired absorbers, but may exclude absorbers that are physically associated but at slightly different redshifts. This could occur for a number of reasons including differing gas velocities or distances at the two probed locations. A larger tolerance will include a larger number of such systems, but may also include chance coincidences of absorbers at similar redshifts that are not truly physically associated. There is no preferred method for choosing an optimized value of the velocity width tolerance $\Delta v_{pair}$, and so we report our results for a number of choices for $\Delta v_{pair}$ as done in previous studies. 

To determine reasonable values of $\Delta v_{pair}$, we compare a normalized histogram of velocity separations between absorbers in sightline pairs of the true data catalog with the same normalized histogram for the simulated catalogs described below. We found that the number of data-data pairs only begins to significantly exceed the number of random-random pairs at $\Delta v_{pair} \leq 500$ km~s$^{-1}$. In other words, there are approximately the same number of data absorbers separated by 500 km~s$^{-1}$ and above as there are random absorbers. To focus on the regime where paired absorbers are less likely to be chance coincidences, we choose to use values for $\Delta v_{pair}$ of 100, 200, and 500 km~s$^{-1}$, but we note that these values differ from those used in \citet{Martin2010}.

Data-data pairs (DD) compare absorbers within the true paired sightlines while random-random pairs (RR) compare absorbers in the simulated catalogs, which have no true physical association. We show a histogram of the DD absorber counts as a function of perpendicular separation of the paired quasars in Figure~\ref{fig:achist}. In this figure, we normalize the counts by the total searchable path in each bin so that a bin with more quasars or with quasars with longer searchable paths will not bias the $r_\perp$ dependence. This normalization removes some of the binning-dependence of the distribution.

We anticipate that quasar pairs with smaller $r_\perp$ (those that are more closely located in the sky) will have a larger number of DD absorber pairs as they will likely probe more of the same galaxy halos than pairs that are farther apart. We see general agreement with this expectation in Figure~\ref{fig:achist}. As described above, for a given $r_\perp$ value, we also see an increasing number of counts per unit searchable path length as $\Delta v_{pair}$ increases from 100 to 500 km~s$^{-1}$ as expected.

\subsection{Random absorbers pair counts}\label{subsec:fakesystems}

We follow the approach of \citet{Martin2010} and run a Monte Carlo simulation using the EW frequency distribution to calculate the expected number of randomly paired absorbers (random-random pairs or RR). Generally, the simulation uses the parameters of the EW frequency distribution function from Section~\ref{sec:distribution} to create 1000 fake \ion{C}{4} catalogs for each quasar that realistically replicate the redshift-dependent detection limitations of the actual spectra but do not contain the physical associations between paired quasars.

We choose to use the EW frequency distribution parameters we derived for the Mintz et al. sample, which as noted above are statistically indistinguishable from those of the Martin et al. sample. In order to integrate and draw samples from the EW frequency distribution function, we must set finite EW limits for the simulation. We choose values of 1 and 1000 m\AA, which represent the EW range of our actual absorber catalog.

To create the simulated catalogs, we first integrate the EW frequency distribution over EW and $dX$ for each sightline to calculate the number of fake absorbers to simulate, setting an upper limit of 50 absorbers per sightline. We then draw an EW for each absorber from the EW frequency distribution using inverse transform sampling and draw a redshift for each absorber from a uniform distribution covering the searchable redshift range for the given spectrum. Each simulated absorber is included in the final catalog if a randomly drawn number is greater than the detection probability for a system of that strength at that redshift in the given spectrum as determined in the sensitivity assessment described in Section~\ref{sec:sensitivity}. Lastly, absorbers in a spectrum closer together than 250 km~s$^{-1}$ were linked as in the compilation of the data catalog.

On average, the resulting simulated catalogs have EW distributions that agree with the true data distribution and have a similar number of systems as the actual data. The average number of systems in each simulated catalog (for all 119 quasars) is 890, compared to 892 systems in the true catalog that have EWs between 1 and 1000 m\AA. These similarities provide evidence that our simulated catalogs accurately reflect the conditions of the spectra.

We calculate the number of RR pairs per sightline pair by averaging over the 1000 runs of the simulation. We show the calculated number of RR pairs binned and normalized by search path as a function of $r_\perp$ with solid bars in Figure~\ref{fig:achist}. As expected, there are significantly fewer RR pairs than DD pairs at all $r_\perp$ and $\Delta v_{pair}$ values.

\subsection{Autocorrelation function}\label{subsec:autocorrelation}

Having calculated the number of DD pairs and the expected number of RR pairs as function of $r_\perp$, we then follow the method used in \citet{Fumagalli2014} and \citet{Scannapieco2006} and calculate the autocorrelation function $\xi$ as: 

$$\xi = \frac{DD}{RR}-1.$$ 

If all of the DD pairs were chance coincidences, we would find DD = RR and therefore $\xi = 0$. A larger value of $\xi$ indicates a higher likelihood of true physical association of DD absorbers. 

While \citet{Martin2010} used a slightly more sophisticated estimator of the autocorrelation function, introduced by \citet{Landy1993} in order to decrease the variance, we choose to use a simpler estimator as our uncertainty will be dominated by Poisson noise due to the small sample size. We calculate the uncertainty on our measurement of the autocorrelation function by propagating the 84.1\% Poisson confidence limits on the absorber pair counts. We show the resulting autocorrelation function in Figure~\ref{fig:ac} for the same values of $\Delta v_{pair}$ reported in Figure~\ref{fig:achist}.

The flattening of the transverse autocorrelation function shown in Figure~\ref{fig:ac} provides a constraint on the minimal extent of \ion{C}{4} in the CGM. A transverse autocorrelation function that is not increasing with decreasing $r_\perp$ indicates that further decreasing the proper separation of the sightlines does not yield significantly more physically associated absorbers. In Figure~\ref{fig:ac}, we see this flattening at $\approx$ 200 $h^{-1}$ comoving kpc for all values of $\Delta v_{pair}$. If these systems had a physical extent significantly smaller than this value, we would expect the autocorrelation function to increase more steeply for smaller $r_\perp$. The flattening we see indicates that quasar sightlines separated by less than $\approx$ 200 $h^{-1}$ comoving kpc are equally likely to probe the same galaxy halo. Provided that the filling factor is sufficiently high, it is possible that the individual clouds of \ion{C}{4} are smaller than this size and we are probing the extent of the \ion{C}{4} gas in the halo. In fact, previous work has shown that individual \ion{C}{4} clouds have physical sizes of at least 300 pc and only begin to exhibit signs of clumping at kpc scales \citep{Rauch2001, Tzanavaris2003}.

While the absolute values of the autocorrelation function are lower than those from \citet{Martin2010} for $\Delta v_{pair}=200$ km~s$^{-1}$, we note that the shape of the function agrees with \citet{Martin2010} for the $r_\perp$ bins covered in that study. As described above, our approach to detect \ion{C}{4} doublets, assess the search sensitivity, and to simulate fake catalogs is not identical to that of \citet{Martin2010}. Each of these differences impacts the number of RR absorbers and therefore the absolute normalization of the autocorrelation function. We note that the absolute normalization of the autocorrelation is largely dependent on the number of RR pairs and the chosen values of $\Delta v_{pair}$. Regardless, our constraint on the extent of \ion{C}{4} is based on the trend of the autocorrelation function which is similar for each of the reported values of $\Delta v_{pair}$, so our conclusions are not dependent on the absolute normalization.

We note that the value at which the flattening occurs is not well specified by our data and is somewhat dependent on the binning of the autocorrelation function. Using a range of 4 – 7 bins, we find that $\xi$ begins to flatten at $\approx$ 200 $h^{-1}$ comoving kpc with a range of $\pm 50$ $h^{-1}$ comoving kpc. While the uncertainties in the autocorrelation are large for $\Delta v_{pair}=200$~km ~s$^{-1}$, they are lower for the other larger $\Delta v_{pair}$ values, and the trend is consistent for all three values of $\Delta v_{pair}$. This consistency over $\Delta v_{pair}$ suggests that the trend is likely real. Furthermore, this value is consistent with other estimates of the spatial extent of \ion{C}{4} from the impact parameters of absorbers to galaxies \citep{Steidel2010, Borthakur:2013, Bordoloi:2014, Rudie:2019, Schroetter:2021, Dutta:2021}.

To better constrain the inflection point of the autocorrelation function from rising to flat, we would require more quasar-pairs with projected separations at $\sim$200 $h^{-1}$ comoving kpc to better sample the $r_\perp$ space. To decrease the uncertainty in $\xi$, we would need more systems per bin, which could be achieved by including more quasar pairs or using spectra with higher SNR, higher resolution, or both. We note, however, that substantially increasing the resolution of the spectra would allow for the detection of significantly weaker \ion{C}{4} systems than is possible with lower resolution data. Such a study would potentially trace a different \ion{C}{4} population outside of the halo. Therefore, the best approach would be to supplement the sample with additional quasar pairs at a similar resolution, none of which are currently available  and so additional data would be required.

\section{Summary}\label{sec:conclusions}
We have conducted a search for \ion{C}{4} absorbers in a sample of high redshift, medium resolution spectra of closely located quasar pairs to investigate the metal distribution and extent of the high redshift CGM. 
\begin{itemize}
    \item We present a new sample of medium resolution spectra of 32 high redshift quasar pairs with redshifts $1.96 \leq z \leq 2.64$ and comoving projected separations of $32 \ h^{-1} \text{comoving kpc} \leq r_\perp \leq 320 \ h^{-1} \text{comoving kpc}$. We supplement our sample with data from \citet{Martin2010} to cover a larger range of redshift and $r_\perp.$
    \item We conduct a doublet search to identify \ion{C}{4} absorbers in the quasar spectra. We find 906 intervening likely absorbers, including 470 from the Mintz et al. sample. The absorbers have equivalent widths $6 \text{ m\AA } \leq W_{r,1550} \leq 2053 \text{ m\AA}$ with a median $W_{r, 1550}$ of 92 m\AA.
    \item We calculate the equivalent width frequency distribution for the \ion{C}{4} 1550 line using the absorber catalog and the results of the completeness assessment. We find that our results agree with the results published in \citet{Martin2010}. We see potential evidence of redshift evolution in the comparison of the two samples, in agreement with predictions from the literature. A larger sample would be necessary for a thorough investigation of the redshift evolution of \ion{C}{4} enrichment or extent, which is impractical with our current set of quasar pairs. 
    \item We compute the autocorrelation function $\xi = \frac{DD}{RR}-1$ as a function of perpendicular separation to assess the significance of the number of paired absorbers found in the quasar pairs. We find that the number of data-data pairs and $\xi$ tend to increase with decreasing $r_\perp$ up to $\approx$~200 $h^{-1}$ comoving kpc, indicating that the halos probed by \ion{C}{4} are likely no smaller than $\approx$~200 $h^{-1}$ comoving kpc.
\end{itemize}

While our results suggest the existence of a lower limit for the physical size of the CGM, we note additional work is needed to better specify this value. To decrease the uncertainty in $\xi$, more systems per bin are required, which would be best achieved by obtaining spectra of additional quasar pairs at similar resolution. To better identify the turnoff point in $\xi$, we would need more quasar pairs at small $r_\perp$ to sample the parameter space more robustly.

Despite the limitations, our results provide a constraint for simulations and demonstrate the power of quasar-pairs and the study of the transverse autocorrelation function as tools for investigating the nature of the CGM. Indeed, the addition of information in the transverse direction provides a valuable metric to constrain the extent of \ion{C}{4} enriched gas around galaxies, beyond what is possible along the line of sight \citep{Scannapieco2006, Martin2010}. As this metric depends on the implementation for feedback, the time at which metals have been injected, and the characteristic mass of the galaxies responsible for the enrichment \citep[e.g.][]{Booth2012}, detailed comparisons of simulations and observations in quasar pairs offer a new way to refine our understanding of metal enrichment in the CGM of galaxies. 

Moreover, by combining measurements on small-scales (e.g. from this work) and larger-scale clustering \citep[e.g.][]{Gontcho2018}, it will be possible to separate the contribution of the one-halo and two-halo terms, to place constraints on both the characteristic halo mass of the galaxies contributing to the \ion{C}{4} absorption and the spatial extent and covering factor of metals in the CGM \citep{Fumagalli2014}. Refined measurements in larger samples together with the analysis of cosmological simulations have the potential to uncover how metals spread and enrich halos at early epochs, beyond what is currently possible by studying absorbers along the line of sight.

\begin{acknowledgements}
This work was supported by NASA Keck PI Data Awards 2016B N032E, 2017A N133E, \& 2019A N027E (PI Rafelski) and 2021B N087 (PI Prichard), administered by the NASA Exoplanet Science Institute. Data presented herein were obtained at the W. M. Keck Observatory from telescope time allocated to the National Aeronautics and Space Administration through the agency's scientific partnership with the California Institute of Technology and the University of California. The Observatory was made possible by the generous financial support of the W. M. Keck Foundation. The authors wish to recognize and acknowledge the very significant cultural role and reverence that the summit of Maunakea has always had within the indigenous Hawaiian community. We are most fortunate to have the opportunity to conduct observations from this mountain. This work is based on observations collected at the European Organisation for Astronomical Research in the Southern Hemisphere under ESO programs PID 099.A$-$0018 and 0100.A$-$0084.This work was supported in part by NSF through AST-1817125 (CLM).
Support for HST Program GO-14127 was provided by NASA through grants from the Space Telescope Science Institute, which is operated by the Association of Universities for Research in Astronomy, Inc., under NASA contract NAS526555. This project was supported in part by the NSF REU grant AST-1757321 and by the Nantucket Maria Mitchell Association. This project has received funding from the European Research Council (ERC) under the European Union’s Horizon 2020 research and innovation program (grant agreement No 757535) and by Fondazione Cariplo (grant No 2018-2329). KHRR acknowledges partial support from NSF grants AST-1715630 and AST-2009417. This research made use of \verb|PypeIt|,\footnote{\url{https://pypeit.readthedocs.io/en/latest/}}
a Python package for semi-automated reduction of astronomical slit-based spectroscopy
\citep{pypeit:joss_pub}.

\end{acknowledgements}

 \facility{} HST (WFC3/UVIS), Keck:II(ESI), VLT(X-Shooter), Magellan(Mage)
 
 \software{ESIRedux (Prochaska et al. 2003), MASE pipeline (Bochanski et al. 2009), linetools (Prochaska et al. 2016), PypeIt (Prochaska et al. 2020)}




\end{document}